\documentclass[a4paper,fleqn]{cas-dc}

\usepackage[authoryear]{natbib}
\usepackage{upgreek}
\newcommand{\mum}{$\upmu$m}
\newcommand{\degree}{$^\circ$}

\begin{document}
\let\WriteBookmarks\relax
\def\floatpagepagefraction{1}
\def\textpagefraction{.001}
\shorttitle{Identification of a new spectral signature at 3~µm over Martian northern high latitudes: implications for surface composition}
\shortauthors{Stcherbinine et~al.}

\title[mode = title]{Identification of a new spectral signature at 3~µm over Martian northern high latitudes: implications for surface composition.}                      



\author[1,2]{Aurélien Stcherbinine}
    [style=french,
    orcid=0000-0002-7086-5443]
\cormark[1]
\ead{aurelien.stcherbinine@ias.u-psud.fr}
\ead[url]{aurelien.stcherbinine.net}

\author[1]{Mathieu Vincendon}
    [style=french]

\author[2]{Franck Montmessin}
    [style=french,
    orcid=0000-0002-4187-1457]

\author[3]{Pierre Beck}
    [style=french]

\address[1]{Institut d'Astrophysique Spatiale, Université Paris-Saclay, CNRS, Orsay, France}
\address[2]{LATMOS / IPSL, UVSQ Université Paris-Saclay, Sorbonne Université, CNRS, Guyancourt, France}
\address[3]{Institut de Planétologie et d'Astrophysique de Grenoble, Université Grenoble Alpes, CNRS, Saint-Martin d'Heres, France}




\cortext[cor1]{Corresponding author}





\begin{abstract}
\noindent 
Mars northern polar latitudes are known to harbor an enhanced 3~\mum\ spectral signature when
observed from orbit. This may indicate a greater amount of surface adsorbed or bound water, although
it has not yet been possible to easily reconcile orbital observations with ground measurements by
Phoenix. Here we re-analyzed OMEGA/Mars Express observations acquired during the Northern summer to
further characterize this 3~\mum\ absorption band increase. We identify the presence of a new
specific spectral signature composed of an additional narrow absorption feature centered at
3.03~\mum\ coupled with an absorption at $\lambda~\geq~3.8~\upmu$m. This signature is homogeneously
distributed over a high-albedo open ring surrounding the circumpolar low-albedo terrains between
$\sim$~68\degree N and 76\degree N and $\sim$~0\degree E and 270\degree E. This location includes
the Phoenix landing site.  This feature shows no time variability and can be confidently attributed
to a seasonally stable surface component. All together, the stability, spectral shape and absence of
significant correlation with other signatures in the 1 -- 2.5~\mum\ range discard interpretations
relying on water ice or easily exchangeable adsorbed water. Sulfates, notably anhydrite, provide
interesting comparisons to several sections of the spectrum. Analogies with Earth samples also show
that the spectral signature could result from a latitudinal modification of the hydration state
and/or grains size of salts contaminants. While the exact full spectral shape cannot be easily
reproduced, plausible explanations to this observation seem to involve geologically recent water
alteration at high northern latitudes.
\end{abstract}




\maketitle

\section{Introduction}  \label{sec:intro}

    Mars ubiquitous wide 3~\mum\ spectral absorption has been first identified using ground
    observations and associated with the presence of hydrated minerals at the surface of the red
    planet \citep{sinton_1967}. This feature was then observed from orbit by e.g. Mariner and Phobos
    missions \citep{pimentel_1974, bibring_1990}. The 3~\mum\ variability over the surface was
    attributed to either difference in absorbed water content, or compositional variations
    \citep{murchie_2000}. The first global mapping of this spectral signature was obtained through
    observations of the Observatoire pour la Minéralogie, l'Eau, les Glaces et l'Activité (OMEGA)
    instrument \citep{jouglet_2007, milliken_2007b}. Either superficial adsorbed water independent
    of composition, bound structural water within minerals, intermediate cases (e.g., water stored
    in the interlayer regions of phyllosilicates), or a mix of these possibilities, were considered
    to explain this feature \citep{milliken_2007b}. Irrelevant of its origin, as
    this spectral feature is mainly associated with OH/H$_2$O, this 3~\mum\ band was used to
    estimate a water weight \% content of the surface \citep{jouglet_2007, milliken_2007b}.
    These studies reveal that the water content of the Martian soil increases poleward by a factor
    greater than two, with the highest levels reached in the northern latitudes ($\geq$ 8 wt. \%
    H$_2$O, up to 10\% to 15\%). While the absolute value of this weight \% is model-dependent
    \citep{jouglet_2007, milliken_2007b, liu_2020}, the relative increase corresponds to a robust
    latitudinal trend of the band depth. This polar increase of hydration was mainly attributed to
    an increase of superficial weakly bound adsorbed water \citep{jouglet_2007, poulet_2008,
    poulet_2010}. This interpretation was notably supported by a tentative identification of
    seasonal variations of the 3~\mum\ band depth, suggestive of exchanges between the surface and
    the atmosphere \citep{jouglet_2007}.
    
    In-situ measurements from the Phoenix lander offered observational "ground truth" constraints
    about the water content of soils at northern polar latitudes. Phoenix reported the detection of
    various minerals that may indicate a past water alteration
    process such as carbonates (2--5\%) and perchlorates ($<$ 1\%), but an overall soil water
    content of 1--2\% only \citep{boynton_2009, hecht_2009, smith_2009a, poulet_2010, sutter_2012} which
    is well below the value of 10-11\% retrieved using OMEGA data at 3~\mum\ \citep{jouglet_2007,
    milliken_2007b, audouard_2014b}. Both measurements were argued to be reconcilable if we consider
    that the 3~\mum\ feature is due to superficial adsorbed water contained within the first few
    hundred micrometers sampled by OMEGA, and not in the first centimeters sampled by Phoenix
    \citep{poulet_2010}. However, water from samples heated by Phoenix was not released at low,
    adsorbed water-compatible temperatures \citep{smith_2009a}. Observed temperatures of
    about 300\degree C and 700\degree C are compatible with the presence of
    several hydrous minerals or phases within soils \citep{smith_2009a}. In the absence of a
    significant amount of adsorbed water, the discrepancy between OMEGA and Phoenix measurements may
    thus be related to either the presence of a very superficial hydrated coating with tightly bound
    - and not adsorbed - water, or to an improper association between 3~\mum\ band depth/shape and
    water \%.
   
    In-situ measurements of the soil water content were also obtained at equatorial latitudes by the
    Mars Sample Laboratory (MSL) rover Curiosity. Various instruments (the Sample Analysis of Mars,
    SAM, the Chemistry and Mineralogy, CheMin, and the Chemistry Camera, ChemCam) lead to the
    conclusion that most of the water (about 2 \%) contained in the martian regolith at MSL landing
    site is likely not weakly bound adsorbed water, but rather more strongly bound water within the
    amorphous phase \citep{leshin_2013, bish_2013, meslin_2013}.  Based on these results,
    \cite{meslin_2013} suggested that the spatial variations of the surface water content observed
    from orbit may be partly related to the variations of abundances of hydrated amorphous
    components.
    
    A reassessment of the 3~\mum\ band in the OMEGA dataset then similarly questioned the
    interpretation of this band by adsorbed water \citep{audouard_2014b}: using a larger dataset and
    applying a stricter atmospheric filtering on the data, these authors did not confirm the
    seasonal variations previously reported from OMEGA data at northern high latitudes. Moreover,
    the surface 3~\mum\ feature was found to be extremely stable with respect to changes in relative
    humidity. It was then suggested that the 3~\mum\ band at high northern latitudes is also not
    linked with adsorbed water, and that the increased 3~\mum\ band may be related to a specific
    polar alteration process involving ice and modifying the amount of tightly bound water within
    minerals or amorphous phases \citep{audouard_2014b}.
    
    Another piece of the puzzle came from laboratory measurements on analogs and meteorites: these
    measurements revealed that increasing the amount of adsorbed water usually creates an absorption
    feature at 1.9~\mum\ in addition to the one at 3~\mum, while this 1.9~\mum\ band is not observed
    over most of Mars unlike the ubiquitous 3~\mum\ band, suggesting again that the Martian 3~\mum\
    band may not be related to adsorbed water \citep{pommerol_2009, pommerol_2011, beck_2015}. Note
    however that under northern polar latitudes, above 60\degree N, the 1.9~\mum\ band is observed
    almost everywhere, with an increasing latitudinal trend similarly to the 3~\mum\ feature
    \citep{poulet_2008}, which is this time consistent with the hydrated water interpretation for
    the 3~\mum\ polar increase \citep{poulet_2008}. As in other context (i.e., at lower latitudes),
    the 1.9~\mum\ band is also an indicator of the presence of hydrated minerals such as
    phyllosilicates \citep{poulet_2005}, the presence of this 1.9~\mum\ may also suggest a specific
    widespread presence of hydrated minerals at high northern latitudes. Actually, the presence of
    hydrated minerals such as phyllosilicates or sulfates also leads to a significant co-increase of
    the 1.9~\mum\ and 3~\mum\ bands \citep[][figure 10]{jouglet_2007}.
    
    Remote observations of the high northern latitudes of Mars indeed reveal the presence of other
    spectral features related to hydrated minerals, in particular sulfates, suggesting the presence
    of water alteration \citep{langevin_2005, poulet_2008, masse_2012, carter_2016}. These
    identifications are based on features located in the 1 to 2.5~\mum\ wavelength range. Near-IR
    data first revealed an area about $\sim$~240\degree E and $\sim$~85\degree N with strong
    signatures compatible with calcium-rich hydrated sulfates such as gypsum \citep{langevin_2005}.
    Note that it has been reported afterward that perchlorates can display similar spectral
    signatures \citep{hanley_2015}. Weaker signatures of these sulfates (and possibly perchlorates)
    have then been observed all over the norther circumpolar dune fields surrounding the perennial
    cap \citep{masse_2010, masse_2012}. Another candidate for some OMEGA hydration features observed
    at high northern latitudes is zeolites \citep{poulet_2010}. Large areas of the northern high
    latitudes actually possess a distinct spectral shape in OMEGA data which may be related to some
    sort of widespread water-limited alteration \citep{horgan_2012}. Such a widespread limited
    weathering potentially producing hydrated phases may also explain data collected by the Thermal
    Emission Spectrometer \citep{wyatt_2004, michalski_2005}.
    
    If hydrated phases concentration within soil is low, such as expected from a limited weathering,
    then they may not be easily identified in remote observations. Some low-abundance minerals that
    have been detected through in-situ measurements at northern high latitudes, as e.g.
    Mg-perchlorates and carbonates at the Phoenix landing site \citep{boynton_2009, hecht_2009} have
    not yet been detected by orbital observations. This may be due to orbital detection threshold
    limitations that can be up to 6\% for carbonates \citep{jouglet_2007b, poulet_2010} or to the
    effect of dust coating which masks underlying materials \citep{vincendon_2015}.
    Similarly, the hydrated amorphous phase observed by Curiosity \citep{leshin_2013, bish_2013,
    meslin_2013} is not identified from orbit. This further suggests that low amount of hydrated
    minerals or amorphous hydrated phases may be widely distributed at high latitudes, without being
    easily identifiable by OMEGA and the Compact Reconnaissance Imaging Spectrometer for Mars
    (CRISM), notably within the most widely used 1 to 2.5~\mum\ spectral range. They could however
    take part in the spectral shape and deepness of the 3~\mum\ H$_2$O/OH feature, which shows up
    more rapidly compared to features in the 1 to 2.5~\mum\ range \citep[e.g.][]{bishop_2019a}. 
    
    Overall, northern latitudes show complex spectral behaviors related to hydrated mineralogy
    and/or adsorbed water, some of which are still not uniquely interpreted nor related to lander
    observations. In this study, we re-analyze OMEGA observations of the
    north polar regions of Mars to further constrain the origin of the polar increase of the 3~\mum\
    feature.

\section{Dataset and methods}   \label{sec:methods}
    \subsection{OMEGA data}     \label{sec:data_reduction}
    The OMEGA experiment onboard the Mars Express orbiter is a visible and near-infrared imaging
    spectrometer composed of three different channels covering the 0.38--5.1~\mum\ spectral range,
    respectively called "V" (0.38--1.05~\mum), "C" (0.93--2.73~\mum) and "L" (2.55--5.1~\mum)
    channels \citep{bibring_2004}.  Observations of the Martian surface using the three channels
    were conducted from 2004 to 2010. Additional observations are still ongoing with limited
    spectral coverage, but in this study, we will only deal with pre-2010 measurements. It
    corresponds to a dataset of 9646 hyperspectral cubes covering most of the Martian surface with a
    typical spatial sampling of 1~km, varying from 350~m to 5~km depending on the position of
    Mars-Express on its elliptical orbit.
    Repeated observations of the same region have been frequently obtained over the mission, in
    particular at high latitudes where time sampling can be about 10\degree\ of $L_s$
    \citep{langevin_2005a,langevin_2007} thanks to the quasi-polar orbit of Mars Express.
    
    While the response of the C channel remains stable from 2004 to 2010, the onboard calibration
    (OBC) of OMEGA reveals that the L channel response has significantly varied over the mission
    \citep{jouglet_2009}.
    In this paper, we use the adapted instrument transfer function (ITF) that accounts for these
    variations, as described in \citet{jouglet_2009}.
    Most of our study will however focus on data acquired during MY 27 northern summer (from orbits
    923 to 1224, i.e. over $L_s$ 98\degree -- 137\degree), a timeframe over which the ITF was
    nominal.
    
\begin{figure}[h!]
        \centering
        \includegraphics[width=\linewidth]{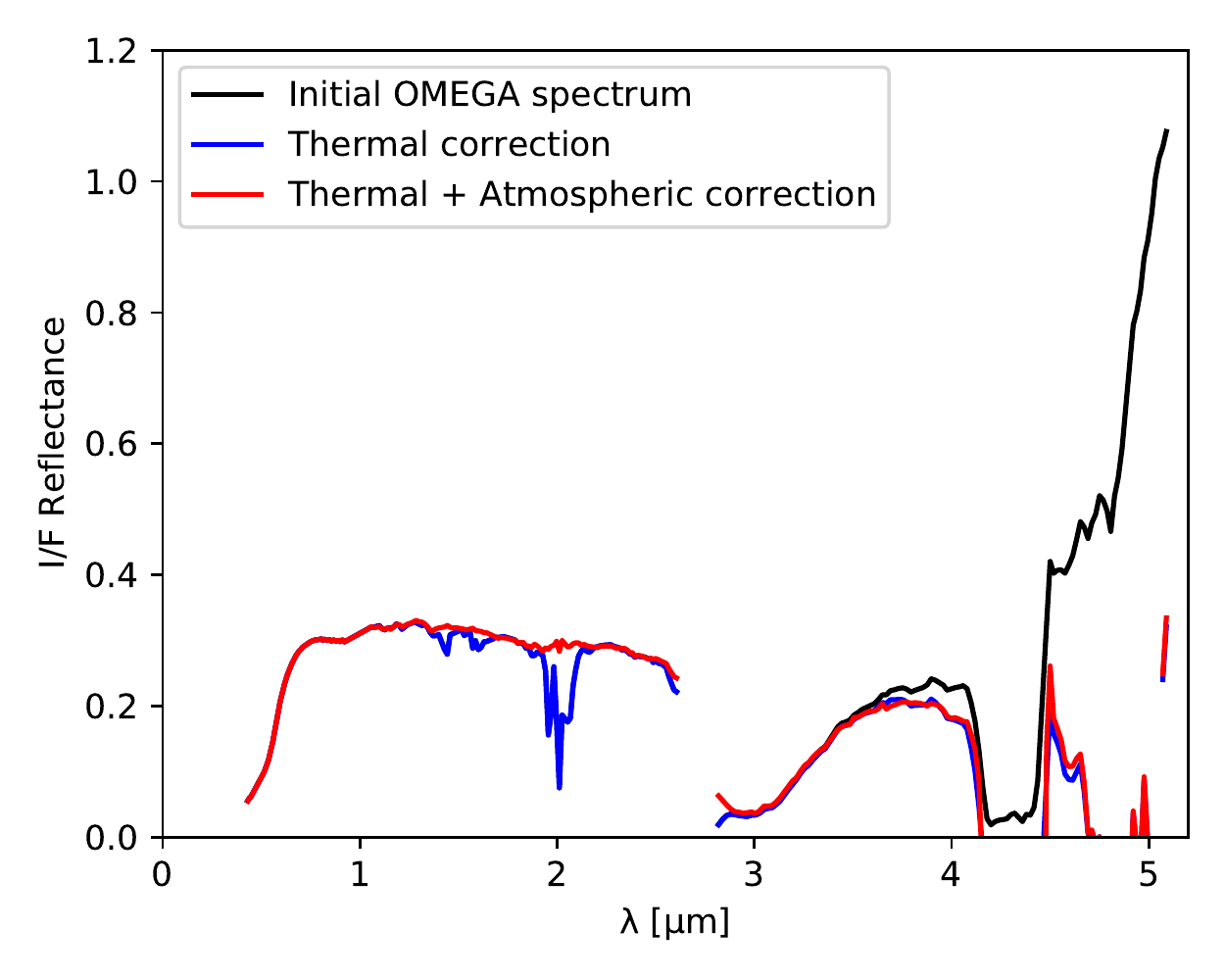}
        \caption{Thermal and atmospheric corrections steps applied to OMEGA spectra in this study.
            The black spectrum is an example of initial OMEGA reflectance spectrum (averaged over 9
            pixels), the blue spectrum is corrected from the thermal component, and the red one is
            also corrected from the atmospheric absorption. See text for details.}
        \label{fig:corrections_therm_atm_sp}
    \end{figure}

    As the thermal emission from the planet is of importance for wavelengths larger than 3~\mum, we
    remove this thermal component from each OMEGA data cube using the method described in
    \citet{jouglet_2007}, and thereafter used in other studies based on OMEGA observations
    \citep[e.g.][]{audouard_2014b}. This method relies on the comparison between the observed
    reflectance at 5~$\upmu$m, where thermal emission dominates for temperatures greater than about
    200~K, and the reflectance at 2.4~$\upmu$m where the thermal contribution is negligible. We
    first estimate the theoretical thermal-free reflectance at 5~$\upmu$m using the measured
    reflectance at 2.4~$\upmu$m, assuming a typical spectral shape extracted from
    \citet{erard_1997}. We then calculate the appropriate thermal additional contribution that
    reproduces the observed reflectance at 5~$\upmu$m. This thermal contribution can then be removed
    at all wavelengths assuming Planck law. An example is shown in
    \autoref{fig:corrections_therm_atm_sp}.
    
    We also correct the OMEGA spectra from the atmospheric gas absorption using the method described
    in \citet{langevin_2005} and \citet{jouglet_2007}. Each OMEGA spectrum is divided by a
    transmission spectrum of the martian atmosphere computed from a comparison between the base and
    the top of Olympus Mons, that is scaled to the appropriated atmospheric column height using the
    2~$\upmu$m atmospheric CO$_2$ absorption band (\autoref{fig:corrections_therm_atm_sp}).
    
    OMEGA data are processed using a Python module developed for this study ("OMEGA-Py"),
    freely available on GitHub at \url{https://github.com/AStcherbinine/omegapy}. This
    module notably includes a re-implementation of the most recent release ("SOFT10") of the
    official OMEGA software (IDL routines) available with OMEGA data. It also contains several
    additional data reduction functions such as build-in atmospheric and thermal corrections
    (using the methods described above), and graphic tools including interactive visualization
    of the data or generation of composite OMEGA maps.
    
    \subsection{Data filtering}     \label{sec:data_filtering}
    OMEGA has provided a large number of observations: we can thus apply strong filtering on the
    dataset while keeping a near global coverage of the polar regions.  This data filtering is based
    on criteria established and detailed in previous studies \citep[e.g.][]{langevin_2007,
    poulet_2007, ody_2012, vincendon_2015} that are summarized below.
    
    \begin{itemize}
        \item To prevent possible effects related to viewing geometry \citep{pommerol_2008a}, we
            restrict our study to near-nadir pixels (emergence angle $<10^\circ$) with an incidence
            angle $<70^\circ$.
        \item As the OMEGA IR detector has been designed and calibrated to operate at temperatures
            below 80~K, we remove the observation lines that have been acquired with higher
            temperatures.
        \item Depending on the hyperspectral cube number, some lines at the beginning and the end 
            of the images have to be removed as they are used for instrument calibration or are
            usually corrupted (see the OMEGA software \emph{SOFT10\_readme.txt} documentation
            file available on the PSA at
            \url{ftp://psa.esac.esa.int/pub/mirror/MARS-EXPRESS/OMEGA/MEX-M-OMEGA-2-EDR-FLIGHT-EXT7-V1.0/SOFTWARE/}
            for more details).
        \item Near-saturation of some pixels can occur when observing a surface element brighter
            than expected (e.g., sun-lighted tilted surfaces).  Similarly to the higher quality
            level criteria defined in \citet{vincendon_2015}, we use a conservative threshold of
            500~DN for the spectel \#40 (corresponding to $\lambda=1.486~\upmu$m), and reject pixels
            with lower value to prevent non-linearity effects.
        \item An instrumental issue perturbs 128 pixels wide observations from orbit 513 to 3283
            (see OMEGA software \emph{SOFT10\_readme.txt} file) as columns \#81 to \#96
            present 44 corrupted wavelengths. Even if it is possible to recover part of the
            information, we decided to remove these columns from the study to focus on the more
            reliable part of the images. 
            Between orbits 2124 and 3283, an additional random noise also affects columns \#68 to
            \#128, which are also removed from our study.
        \item The 3~\mum\ band is very sensitive to the presence of water ice at the surface or in
            the atmosphere \citep{vincendon_2011}. We remove all pixels with a water ice 1.5~\mum\
            absorption (see \autoref{tab:spectral_estimators}) $>1\%$.  We also compute the ice
            cloud index (ICI) defined in \citet{langevin_2007} as the ratio of the reflectance at
            $\lambda=3.40~\upmu$m over the reflectance at $\lambda=3.53~\upmu$m to detect the
            presence of clouds and remove pixels with an ICI $<0.8$ \citep{audouard_2014b}.
    \end{itemize}
    
    In addition, 3~\mum\ band depth estimators are known to exhibit a strong dependency to the
    surface albedo \citep{jouglet_2007,pommerol_2008}, so a specific attention should be taken when
    comparing regions of different albedo.  Thus, in the following, we usually apply restrictions on
    the value of the reflectance at $\lambda=2.26~\upmu$m when studying the spatial evolution of a
    given criterion.

    \subsection{Band depth estimators}      \label{sec:spectral_estimators}
        
        \begin{table*}[h!]
            \centering
            \begin{tabular}{lc}
                \toprule
                Description & Spectral indice \\
                \midrule
                1.5~\mum\ BD \citep{poulet_2007} & 
                    $\displaystyle 1 - \frac{R(1.50) + R(1.51)}{R(1.30) + R(1.71)} $\\
                1.9~\mum\ BD \citep{langevin_2005a} & 
                    $\displaystyle 1 - \frac{R(1.93)}{0.74 R(1.86) + 0.26 R(2.14)} $\\
                2.4~\mum\ BD & 
                    $\displaystyle 1 - \frac{\overline{R}(2.43, 2.45, 2.46)}
                    {\overline{R}(2.27, 2.29, 2.30)}$\\
                \multirow{2}{*}{\emph{Wide} 3~\mum\ BD \citep{jouglet_2007}} & 
                    $\displaystyle 1 - \frac{1}{N_\lambda} \sum_{\lambda = 2.9\mu m}^{3.7\mu m}
                    \frac{R(\lambda)}{Cont(\lambda)}$ \\
                    & where
                    $\displaystyle Cont(\lambda) = R(2.35) + \left(\frac{R(3.7) - R(2.35)}{1.35}\right)
                    \left(\lambda - 2.35\right) $ \\
                \emph{Narrow} 3~\mum\ BD & 
                    $\displaystyle 1 - \frac{\overline{R}(2.96, 3.01, 3.03)}
                    {0.67\overline{R}(2.88, 2.90, 2.92) + 0.33\overline{R}(3.19, 3.21, 3.23)}$\\
                4~\mum\ BD & 
                    $\displaystyle 1 - \frac{\overline{R}(3.94, 3.96)}
                    {\overline{R}(3.69, 3.71, 3.73)}$\\
                \bottomrule
            \end{tabular}
            \caption{Spectral estimators used in this study.  $R(\lambda)$ is the reflectance
                I/F$\cos(i)$ (corrected from both atmospheric and thermal contributions) at the
                given wavelength $\lambda$ in \mum, and $\overline{R}(\lambda_1, ..., \lambda_n)$ is
                the average reflectance from the $n$ spectels corresponding to wavelengths
                $\lambda_1, ..., \lambda_n$.
                For the wide 3~\mum\ BD, $Cont(\lambda)$ is the used spectral continuum defined
                linearly between 2.35~\mum\ and 3.7~\mum, and $N_\lambda$ is the number of spectels
                between 2.9~\mum\ and 3.7~\mum.}
            \label{tab:spectral_estimators}
        \end{table*}
        
        The spectral estimators used and developed in this study are summarized in
        \autoref{tab:spectral_estimators}, with their explicit formula.
        
        We first consider three band depths estimators derived from previous studies at
        1.5~\mum\ (water ice), 1.9~\mum\ (hydrated minerals or adsorbed water) and 3~\mum. For the
        1.5~\mum\ and 1.9~\mum\ band depth (BD), we compute the absorption at the center of the band
        relatively to a continuum defined linearly between two reference points at the extremities
        of the band. Wavelengths are selected to avoid atmospheric absorption (see
        \autoref{sec:data_reduction}). E.g., this is why the reflectance at 1.93~\mum\ is used in
        the computation of the 1.9~\mum\ BD instead of the reflectance at 1.94~\mum\
        \citep{langevin_2005}. For the quantification of the absorption related to the full 3~\mum\
        band observed everywhere on Mars, we have computed the 3~\mum\ normalized integrated band
        depth under a continuum defined linearly between 2.35~\mum\ and 3.7~\mum, as described in
        \citet{jouglet_2007}. This band depth is hereafter referred as the \emph{wide} 3~\mum\ BD.
        
        We have then defined three new estimators which appeared necessary during the development of
        this study. A new estimator focus on the central part of the 3~\mum\ band: it makes it
        possible to locate a narrow sub-feature within the wide 3~\mum\ band. It is called hereafter
        the \emph{narrow} 3~\mum\ BD: the absorption is computed at 3.0~\mum\ relatively
        to a continuum whose reference points are taken at 2.9 and 3.2~\mum\ instead of 2.9
        and 3.7~\mum\ for the usual, previously used wide 3~\mum\ BD. Indeed, as the usual wide
        3~\mum\ feature is almost linear between 2.9 and 3.2~\mum, this criterion traces
        the concavity in this spectral range related to the presence of the narrow 3~\mum\ band
        (cf. \autoref{fig:saut_phoenix}b). As most wavelengths used for this
        criterion are localized around the bottom of the 3~\mum\ band, reflectance values are low
        due to the strong absorption, and we average 3 consecutive wavelengths (see 
        \autoref{tab:spectral_estimators}) to increase the signal-to-noise ratio (SNR).
        
        Another spectral feature identified during this work is located at about 4~\mum. In the case
        of this 4~\mum\ band, as we can only see one part of the band because of the strong
        atmospheric absorption for wavelength larger than 4.06~\mum, we compute a new 4~\mum\ BD
        estimator using a linear continuum equals to the average reflectance value at $\sim$
        3.7~\mum.
        
        Finally, we will also explore possible correlations with another minor band at 2.4~\mum\ for
        which a new estimator has also been developed, using the calculation principles of the
        4~\mum\ BD (\autoref{tab:spectral_estimators}).

\section{Results}   \label{sec:results}
    \subsection{Phoenix landing site area}     \label{sec:region_phoenix}
        As discussed in \autoref{sec:intro}, Phoenix provided unique ground truth measurements in
        the north polar region that can be compared to information derived from orbital data.
        
\begin{figure*}
            \centering
            \includegraphics[width=\textwidth]{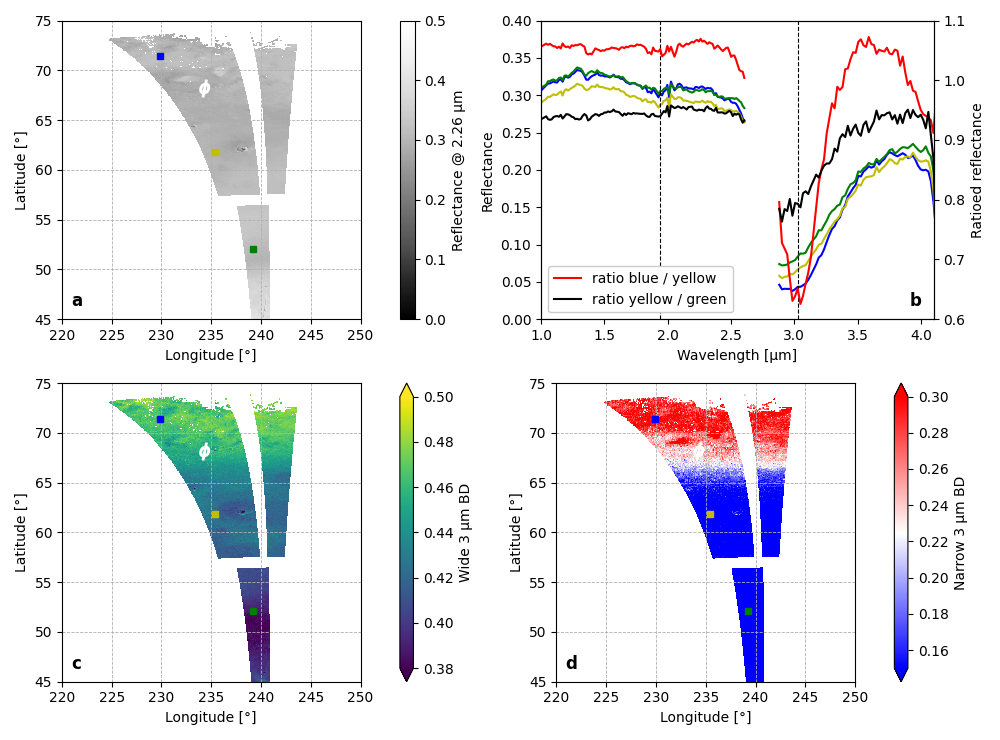}
            \caption{Maps of OMEGA cubes \#3 \& \#4 from orbit 979 ($L_s$ = 105\degree, local time
                $\sim$ 3~p.m.) showing the surface reflectance at 2.26~\mum\ (a), the wide 3~\mum\
                BD (c) and the narrow 3~\mum\ BD (d).  The white "$\phi$" indicates the position of
                the Phoenix lander.
                The blue, yellow, and green spectra of the panel (b) are averaged on a square of $3
                \times 3$ pixels whose position is indicated by the colored squares on
                panels (a), (c) \& (d). Two ratios of these spectra are also shown on panel (b) in red and black.
                Dotted lines on panel (b) correspond to wavelengths of 1.94 and 3.03~\mum.  We observe on
                panel (b) that the evolution between 60\degree N and 70\degree N (red ratio) is very
                different from the one between 50\degree N and 60\degree N (black ratio) with a
                narrow signature centered around 3~\mum, which results in a shift
                of the position of the maximum of absorption of the widespread Martian 3~\mum\ band
                from 2.9~\mum\ (yellow \& green lines) to 3.0~\mum\ (blue line). 
                We also observe an absorption at $\lambda~>$ 3.6~\mum\ specific to the red ratio and
                associated blue spectrum.}
            \label{fig:saut_phoenix}
        \end{figure*}

        \autoref{fig:saut_phoenix} illustrates the evolution of surface spectral reflectance
        over a latitudinal band covering a relatively uniform bright albedo area and crossing the
        Phoenix landing site at 234\degree E / 68\degree N. We can see that the wide 3~\mum\ BD
        progressively increases with latitude, as previously reported \citep{jouglet_2007,
        milliken_2007b}. In panel (b) of the same figure, we show spectra from three different
        latitudes (52\degree N, 62\degree N \& 72\degree N). We can see that the spectral
        modifications are not progressive with latitudes, in particular between 2.8 and 4.0~\mum.
        Indeed, we observe that the shape of the northern spectrum differs from the two others (blue
        versus yellow and green on \autoref{fig:saut_phoenix}): the maximum of absorption of
        the 3~\mum\ band is shifted from $\sim$ 2.9~\mum\ to $\sim$ 3.0~\mum, and a decrease of the
        reflectance is observed beyond $\sim$ 3.8~\mum. This difference of behavior is enhanced in
        the spectral ratios (\autoref{fig:saut_phoenix}b): while the ratio obtained south of
        Phoenix (black) reveals an increase of the 3~\mum\ band with a typical band shape similar to
        the actual band prior ratioing (minimum at $\lambda \sim$ $<$ 2.9~\mum), the ratio
        obtained poleward, over the Phoenix area (red), exhibits a strong but narrow absorption band
        centered on $\lambda \sim 3.03~\upmu$m (hereafter referred as the \emph{narrow} 3~\mum\
        band) along with a wide shallow absorption band starting at $\lambda \sim 3.6~\upmu$m.
        
        We have then mapped the spatial distribution of this narrow 3~\mum\ band in the surroundings
        of Phoenix using the new associated spectral criteria defined in
        \autoref{tab:spectral_estimators} and \autoref{sec:spectral_estimators}. This mapping
        (\autoref{fig:saut_phoenix}d) reveals that this band is observed only above 68\degree
        N. The transition between the area without signature ($<$ 66\degree N) towards that with the
        signature ($>$ 68\degree N) is abrupt, which is very different from the progressive
        latitudinal gradient observed for the wide 3~\mum\ band (\autoref{fig:saut_phoenix}c).
        
        \subsection{Spatial distribution}    \label{sec:north_regions}
\begin{figure*}
            \centering
            \includegraphics[width=\textwidth]{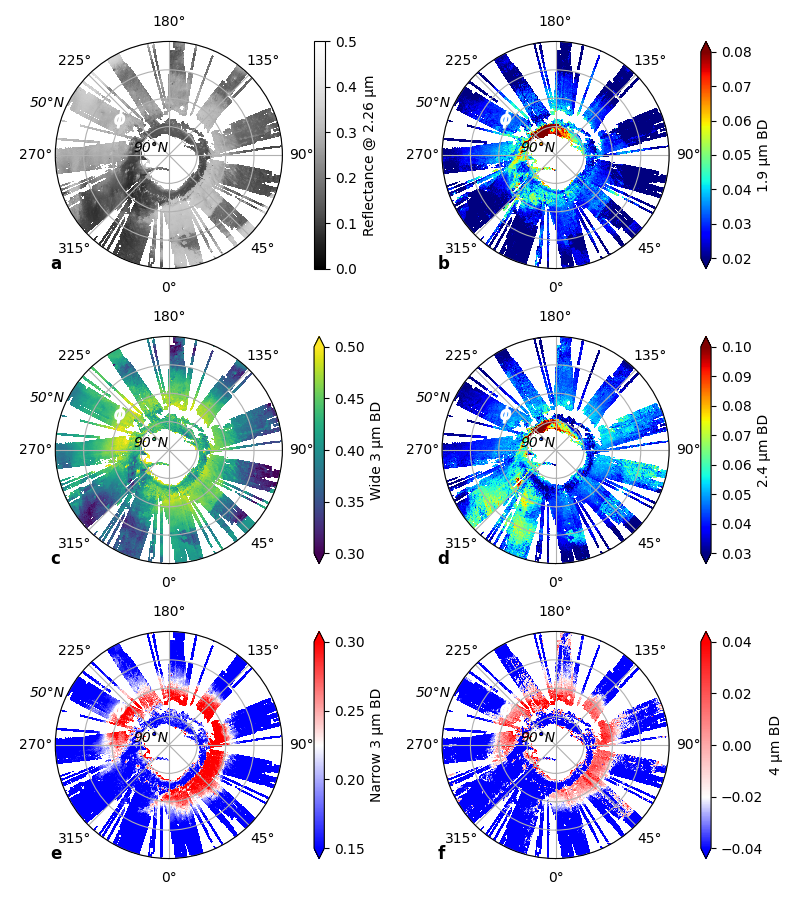}
            \caption{Maps of the martian north polar regions, showing the surface reflectance at
                2.26~\mum\ (a), the 1.9~\mum\ BD (b) the wide 3~\mum\ BD (c), the 2.4~\mum\ BD (d),
                the narrow 3~\mum\ BD (e) and the 4~\mum\ BD (f).
                The white "$\phi$" indicates the position of the Phoenix lander.
                These composite maps are assembled from OMEGA nadir observations during the northern
                summer of MY 27 ($L_s=98^\circ - 132^\circ$), when the response of the C-channel is
                nominal, i.e. from orbit 923 to 1223 \citep{jouglet_2009}.\\
                We observe on panels (e) and (f) the presence of an open ring structure between
                $68^\circ$N and $76^\circ$N (referred to the \emph{3~$\mu$m northern
                ring}): this surface distribution is specific to the narrow 3~\mum\ BD and 4~\mum\
                BD criteria. This region is not correlated with surface albedo variations although
                it occurs only in bright albedo terrains (a). It is nearly anti-correlated with the
                1.9~\mum\ (b) and 2.4~\mum\ (d) absorption features that tend to be observed mainly
                in dark terrains. The wide 3~\mum\ BD is also shown for comparison in panel (c).}
            \label{fig:multi_polar_maps_nord}
        \end{figure*}

\begin{figure}[h!]
            \centering
            \includegraphics[width=\linewidth]{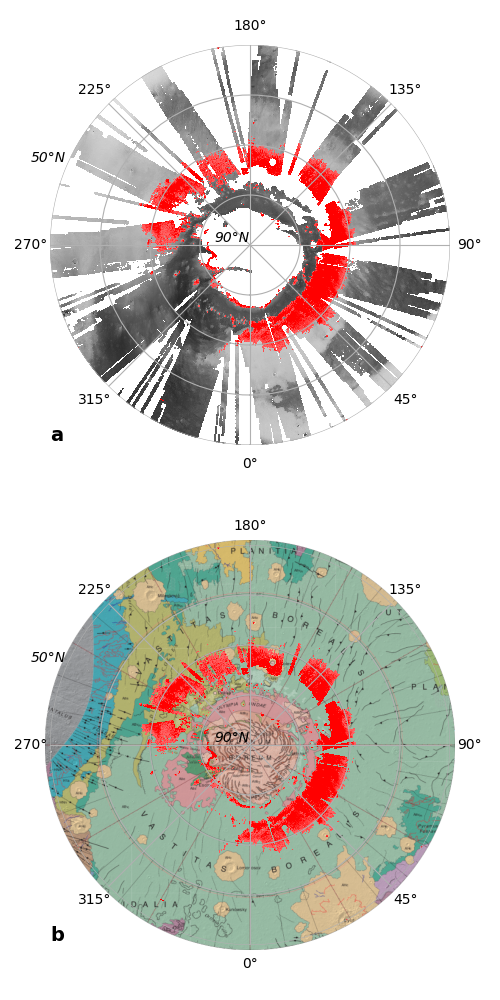}
            \caption{Overplot of the 3~\mum\ northern ring, defined as the region displaying a
                narrow 3~\mum\ BD larger than 0.27 (red) over a map of the surface reflectance at
                $\lambda=2.26$~\mum\ (a), and a geological map of the Martian north regions from
                \citet{tanaka_2005} (b). We observe that the narrow 3~\mum\ signature is
                distributed over two main geological units dated Early Amazonian (see
                \citet{tanaka_2005} for the geological units legend).}
            \label{fig:geol_map_saut}
        \end{figure}

        Outside the Phoenix landing site region, we computed the narrow 3~\mum\ BD over the whole
        OMEGA dataset that contains observations with the 3 channels operating (i.e. up to orbit
        8485 and the loss of the C channel) to map the spatial distribution of the newly identified
        signature at $\lambda \sim 3.03~\upmu$m. It turns out that the signature is only present in
        the north polar regions. \autoref{fig:multi_polar_maps_nord} shows composite maps of
        the northern hemisphere (down to 50\degree N), and we observe on panels (e) and (f) that the
        narrow 3~\mum\ band is spread over an open ring, crescent-shaped area around the North pole.
        This open ring area, hereafter referred as the \emph{3~$\mu$m northern ring} or \emph{ring},
        corresponds to bright regions at latitudes comprised between $\sim$ 68\degree N and
        76\degree N and longitudes between $\sim$~0\degree E and 270\degree E. The lower (southern)
        transition to the ring can occur in bright terrains, without changes in terms of surface
        albedo. Between 30\degree E and 115\degree E, the transition is located close to the
        frontier with low-albedo terrains. The upper (northern) limit of the ring (around 76\degree
        N) corresponds to the delimitation between the high-albedo northern terrains and the dark
        regions around the perennial north polar cap (cf \autoref{fig:geol_map_saut}a).
        Similarly, no detection has been observed in the southward extension of this low-albedo
        region in Vastitas Borealis, between 280\degree E and 360\degree E, while latitudes are
        comparable to the 3~\mum\ northern ring area. The new detection thus seems to require two
        conditions for observation: bright albedo and latitudes greater than $\sim$ 68\degree N.
        
        \begin{figure*}[p]
            \centering 
            \includegraphics[width=\textwidth]{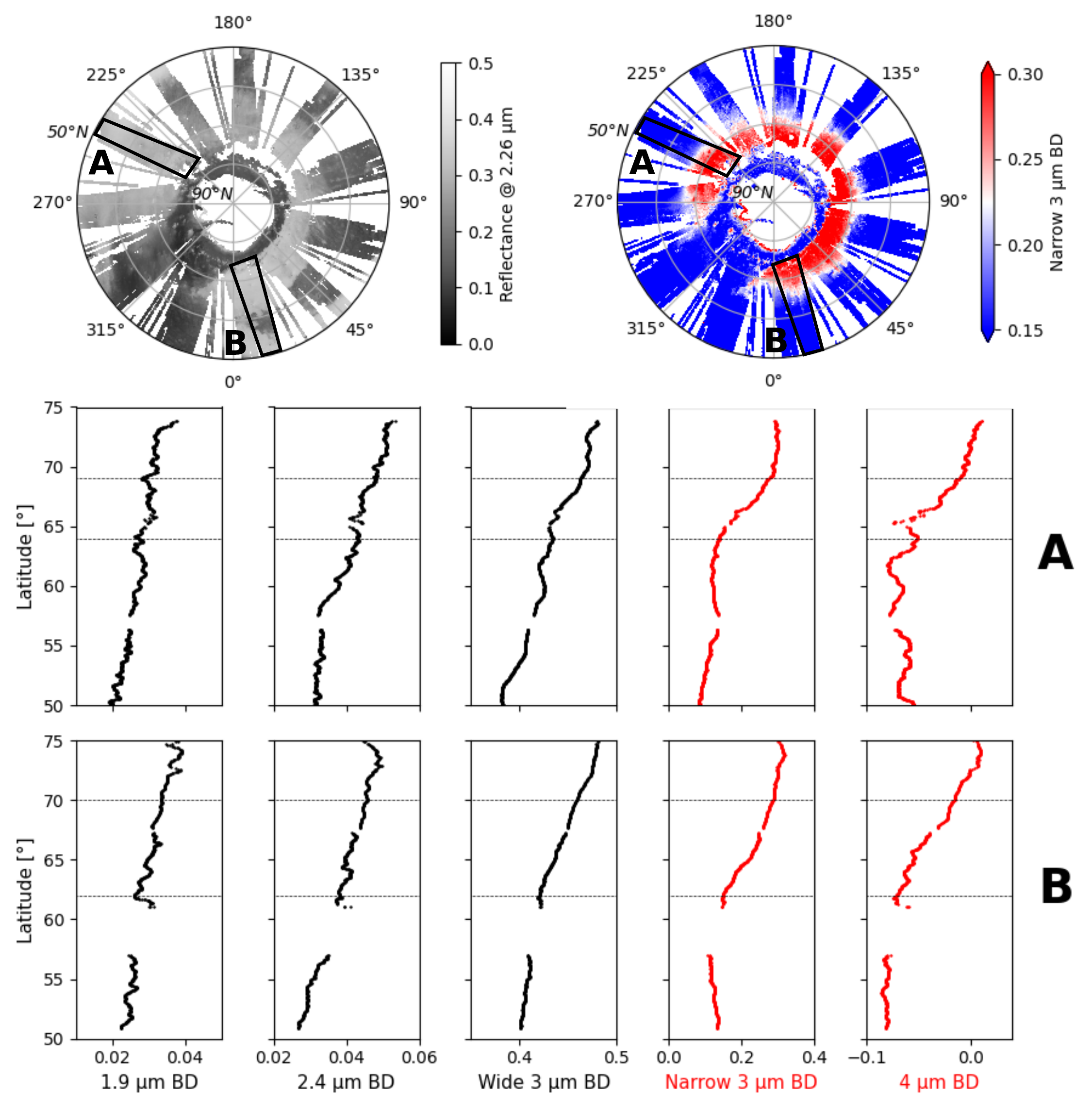}
            \caption{Latitudinal variations of the 1.9~\mum\ BD, 2.24~\mum\ BD, wide 3~\mum\ BD,
                narrow 3~\mum\ BD and 4~\mum\ BD between 50\degree N and 75\degree N, from two
                bright regions identified on the albedo and narrow 3~\mum\ BD maps.
                A: OMEGA cubes 3 \& 4 of orbit 979, $L_s=105^\circ$, MY 27, local time $\sim$ 15h,
                longitude $\sim$ 230\degree E -- 243\degree E (see \autoref{fig:saut_phoenix}).
                B: OMEGA cubes 2 \& 3 of orbit 941, $L_s=100^\circ$, MY 27, local time $\sim$ 15h,
                longitude $\sim$ 5\degree W -- 20\degree E.
                Each point is averaged in longitude (image lines), considering only the pixels with
                a surface reflectance at 2.26~\mum\ between 0.3 and 0.35 to avoid albedo effects in
                the latitudinal trends of the band depth estimators. Profiles are then smoothed
                using a moving average.  We observe on panel A that if the 1.9~\mum, 2.4~\mum\ and
                wide 3~\mum\ bands show a continuous increase between 50\degree N and 75\degree N
                (black lines), the narrow 3~\mum\ band behaves differently with a step profile: the
                narrow 3~\mum\ BD $\sim$ is at 0.12 below 64\degree N, and jump to $\sim$ 0.3 above
                69\degree N. And a similar two-component trend is also observed for the 4~\mum\
                band.
                For the region of panel B, we observe a less sharp evolution between the two narrow
                3~\mum\ band populations, as the transition between the two regimes occurs from
                $\sim$ 62\degree N to $\sim$ 70\degree N.}
            \label{fig:profils_lat}
        \end{figure*}
        
        \autoref{fig:geol_map_saut} shows the spatial distribution of the detection (narrow 3~\mum\
        BD values larger than 0.27) over a map of geological units from \citet{tanaka_2005}. We
        observe that it is spread across two main Early Amazonian geological units (referred as
        Vastitas Borealis marginal and interior units in \citet{tanaka_2005}). This suggests that
        the signature is associated with the superficial composition of the soil overlying both
        units (OMEGA sensing depth is a few hundreds of micrometers only). 
    
        We show in \autoref{fig:profils_lat} latitudinal profiles of band depth averages between
        45\degree N and 75\degree N for two areas crossing the 3~\mum\ northern ring ("A" and "B").
        Four spectral estimators are shown: wide 3~\mum\ BD, 1.9~\mum\ BD, narrow 3~\mum\ BD, and
        4~\mum\ BD. The values of spectral estimators are averaged over longitude (i.e. image
        lines), considering only the pixels with a surface reflectance at 2.26~\mum\ between 0.3 and
        0.35 to avoid possible interference due to albedo variations.  We observe that the wide
        3~\mum\ BD and 1.9~\mum\ BD continuously increase with the latitude. On the other hand, both
        the narrow 3~\mum\ and 4~\mum\ bands present an abrupt increasing step from low to higher
        values over a few degrees of latitude only.
        
        Actually, as shown in \autoref{fig:multi_polar_maps_nord}, the spatial distribution
        of the 4~\mum\ band is similar to that of the narrow 3~\mum\ one. We can see that the
        correlation between both is excellent. This strongly suggests that both are connected, i.e.
        that the phenomena ongoing over the 3~\mum\ northern ring is responsible for both absorption
        features.

        \subsection{Correlation with other spectral features}   \label{sec:spectres_moyens}
        We have averaged several latitudinal spectral ratios ($\sim$ 70\degree N over $\sim$
        63\degree N) extracted all around the 3~\mum\ northern ring to isolate the mean spectral
        signature of this newly detected area (see red spectrum in
        \autoref{fig:spectres_typiques}).
        
        We first compare this signature with another spectral ratio taken south of the ring
        ($\sim$60\degree N over $\sim$50\degree N). This ratio shows that the 3~\mum\ increase from
        $\sim$50\degree N to $\sim$60\degree N has a different spectral shape compared to that
        observed at northern latitudes over the ring. We observe that the main differences between
        these two spectral ratios reside in the 2.8 -- 4~\mum\ spectral range. The increase of
        absorption at 3~\mum\ is noticeably larger between 63\degree N and 70\degree N than between
        50\degree N and 60\degree N, while the latitudinal range is smaller, suggesting an
        enhancement of the 3~\mum\ increase over the ring.  As previously discussed for the Phoenix
        region (\autoref{sec:region_phoenix}), we observe again that the spectral shape between both
        also significantly differs. Over the 3~\mum\ northern ring, we observe a narrow absorption
        centered on 3.03~\mum\ along with what may be the beginning of a wide absorption at
        wavelengths larger than 4~\mum\ (starting at $\lambda \sim$ 3.6~\mum). Southward, we observe
        a wider 3~\mum\ absorption centered at $\lambda \sim$ 2.9~\mum\ and no 4~\mum\ absorption
        (cf \autoref{fig:spectres_typiques}c).
        
        Having a closer look at the 1 -- 2.6~\mum\ spectral range, we can observe the presence of
        some very small absorption features (typically about one percent of absorption in the
        ratios).  The main features that can be identified in the red spectrum are localized at
        1.37, 1.9, 2.4 and 2.6~\mum. The 1.37~\mum\ and 2.6~\mum\ band are associated with
        atmospheric water vapor \citep{encrenaz_2005}. As we compute spectral ratio across about ten
        degrees of latitude, the presence of these water vapor bands in the spectra is consistent
        with the latitudinal increase of the water vapor content characteristic of this season
        during which water vapor is released from the cap  \citep[e.g.][]{smith_2006,
        melchiorri_2007, trokhimovskiy_2015}. Note also that the solar incidence angle increases as
        we go poleward, which results in longer atmospheric path lengths that strengthens
        differences in ratios.  The 1.9~\mum\ and 2.4~\mum\ bands are observed for various hydrated
        minerals as they result from the stretching and bending vibrations of H$_\textrm{2}$O
        molecules and -OH groups \citep{gendrin_2005a, bishop_2019a}.
        However, the mapping of these absorption bands in the North polar regions
        (\autoref{fig:multi_polar_maps_nord}b\&d) and their latitudinal profiles in the bright
        regions (\autoref{fig:profils_lat}) show that their intensity tends to increase with
        latitude, but without being correlated to the 3~\mum\ northern ring.  In addition, regions
        that exhibit the strongest absorption at 1.9~\mum\ and 2.4~\mum\ are not located in the ring
        either (see \autoref{fig:multi_polar_maps_nord} and also \citet{poulet_2008}).

        \subsection{Comparison with the Southern hemisphere}   \label{sec:southern_hemisphere}
\begin{figure}[h!]
            \centering
            \includegraphics[width=.87\linewidth]{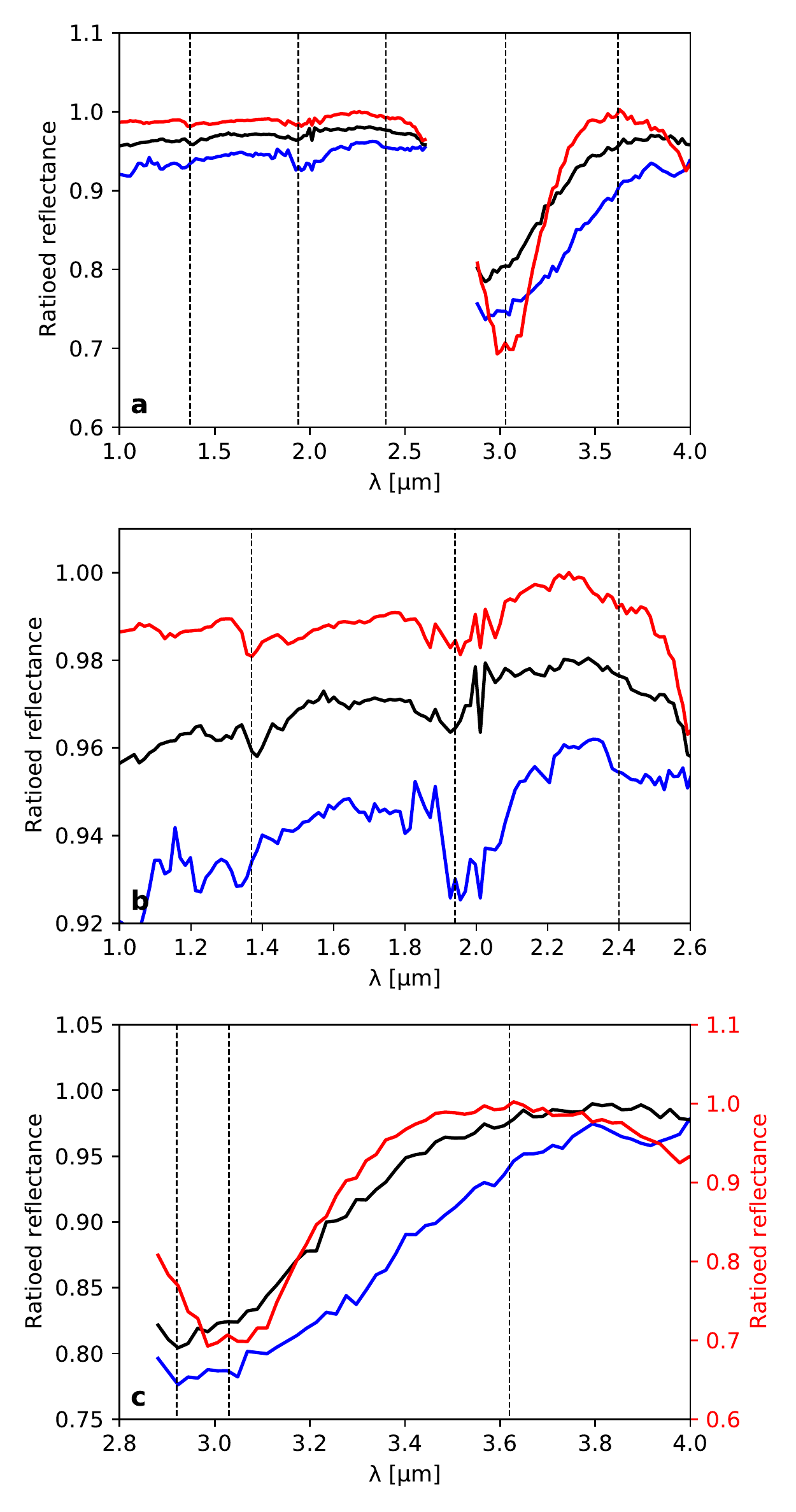}  
            \caption{(a) A typical averaged spectral ratio corresponding to the 3~\mum\ northern
                ring identified in this study (red) is compared to a ratio obtained outside the ring
                between 50\degree N and 60\degree N (black), and to a ratio from southern bright
                terrains between 62\degree S and 74\degree S (blue). 
                Black dotted lines correspond to wavelengths of 1.37, 1.94, 2.4, 3.03, and
                3.62~\mum.  The red spectrum is the average of 4 ratioed spectra taken along the
                narrow 3~\mum\ northern ring (longitudes $\sim$ 15\degree E / 135\degree E /
                240\degree E / 265\degree E). 
                Each one is the ratio of the average of 200 spectra in a bright region at latitude
                $\sim$70\degree N over a similar region in albedo at the same longitude and at
                latitude $\sim$63\degree N, then normalized to the reflectance at
                2.26~\mum\ before averaging ratios from different longitudes.
                The black spectrum is the average of 3 spectra acquired at different longitudes
                ($\sim$ 15\degree E / 240\degree E / 265\degree E). Each one is the ratio of 200
                pixels from a bright region at $\sim$60\degree N over a region with a similar albedo
                at the same longitude at $\sim$50\degree N.
                The blue spectrum is the average of 2 spectra acquired at different longitudes
                ($\sim$ 140\degree E / 205\degree E). Each one is the ratio of 200 pixels from a
                bright region at $\sim$74\degree N over a region with a similar albedo at the same
                longitude at $\sim$62\degree N.
                (b) \& (c) Zooms on the signatures of the C and L channels respectively.
                On panels (a) \& (b) the blue and black spectra have been offset for clarity.  On
                the panel (c) the y-scale of the red spectrum has been adjusted independently for
                clarity.}
            \label{fig:spectres_typiques}
        \end{figure}

        \autoref{fig:spectres_typiques} also presents a typical ratio from the southern polar
        regions (blue spectrum). As no narrow 3~\mum\ band has been detected in the Southern
        hemisphere during mapping, the location of the used spectra to calculate the ratio ($\sim$
        74\degree S over $\sim$ 62\degree S, blue spectrum) were chosen to be as close as possible
        to the conditions of observation of the signature in the Northern hemisphere in terms of
        albedo and latitude. This spectral ratio confirms that the spectral band shape observed over
        the 3~\mum\ northern ring is specific to the northern high latitudes: we can see that the
        maximum of absorption of the 3~\mum\ band occurs at $\sim$ 2.9~\mum\ as for the black
        spectrum, and that there is no absorption at 4~\mum. 
        
        We can notice that several absorption features are present in the 1 -- 2.6~\mum\ spectral
        range in the southern hemisphere, and that they also differ from the north. This will not be
        discussed here, but information can be found in \cite{poulet_2008} and \cite{carter_2016}.

        \subsection{Time stability}     \label{sec:temporal_stability}
        We have looked for time variations of this signature, as a function of year,
        season, day, or hour.
        
\begin{figure}[h!]
            \centering
            \includegraphics[width=\linewidth]{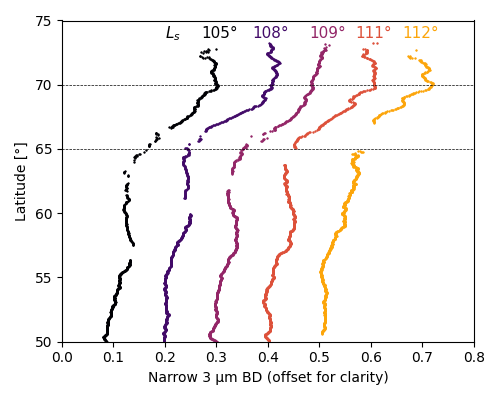}
            \caption{Profiles of the latitudinal evolution of the narrow 3~\mum\ BD (offset for
                clarity) from several observations of the Phoenix landing site region at different
                $L_s$ (orbits 0979, 1001, 1012, 1023 \& 1034).
                Each point is averaged in longitude between 235\degree E and 240\degree E (image
                lines), considering only the pixels with a surface reflectance at 2.26~\mum\ between
                0.3 and 0.35 to avoid albedo effects in the latitudinal trends of the band depth
                estimators.  Profiles are then smoothed using a moving average.}
            \label{fig:compare_profiles_ls}
        \end{figure}

        Due to the Mars Express orbit and our restrictions applied on the emergence and incidence
        angle (cf \autoref{sec:data_filtering}), all the observations used here to compute maps from
        the Martian northern summer of MY 27 have been acquired in the middle of the afternoon
        (around 3~p.m.).  \autoref{fig:compare_profiles_ls} shows latitude profiles above the
        Phoenix landing region derived with 5 successive OMEGA observations taken over $\sim$ 15
        days ($L_s$ 105\degree -- 112\degree).  We can see that apart from some small variations due
        to longitudinal spatial discrepancies between the observations, the narrow 3~\mum\ BD
        profile remains stable.
        
        We have also compared OMEGA observations acquired in the morning ($\sim$~9~a.m.) and evening
        ($\sim$~6~p.m.) over different Martian Years (MY 28 to 30). Morning and evening observations
        imply differences in the observation geometry, essentially in terms of incidence angle:
        comparing both is not trivial. We did not identify significant variation in the narrow
        3~\mum\ BD criteria, despite a wide amplitude in parameters varying with time such as
        surface temperature.
        
        Overall, available observational constraints suggest that the spatial distribution and
        spectral shape of this signature are stable with time.
        
\section{Discussion}
    \subsection{Potential atmospheric biases}     \label{sec:discussion_atm}
    OMEGA observations of the surface are obtained through the atmosphere. Atmospheric constituents
    could create an absorption at 3~\mum\ or a modification of the 3~\mum\ band depth or shape.
    There are three possible atmospheric contributors in this wavelength range: water vapor, water
    ice clouds/fog, and dust.
    
    Water vapor concentration is at its highest value during northern summer close to the polar cap,
    with a latitudinal gradient \citep{smith_2004,smith_2006}. A spectral ratio between high and low
    latitudes could thus highlight more water vapor, as discussed in
    \autoref{sec:spectral_estimators}. However, the impact of water vapor on the 3~\mum\ band was
    previously shown to be negligible considering the expected depth of water vapor features
    \citep{jouglet_2007}. 
    
    Atmospheric dust is known to modify significantly the depth of the 3~\mum\ feature
    \citep{audouard_2014b}. Dust is however not associated with a narrow 3~\mum\ spectral feature
    such as observed here \citep{vincendon_2009}. Moreover, the amount of dust in the atmosphere,
    its variability, and its impact on surface spectra has been studied in details for the MY27
    northern summer OMEGA observations of the high latitudes \citep{vincendon_2007}: the optical
    depth of dust decreases by a factor of 2-3 from the summer solstice to mid-summer to reach a low
    value of about 0.3 for which we do not expect any major impact on the 3~\mum\ band
    \citep{audouard_2014b}. Additionally, we do not observe any major modification of the detected
    spectral feature over that period of changing optical depth (\autoref{fig:compare_profiles_ls}).
    Overall, it thus seems highly unlikely that the observed spectral feature over the ring can be
    explained by atmospheric dust.
    
\begin{figure}[h!]
        \centering
        \includegraphics[width=\linewidth]{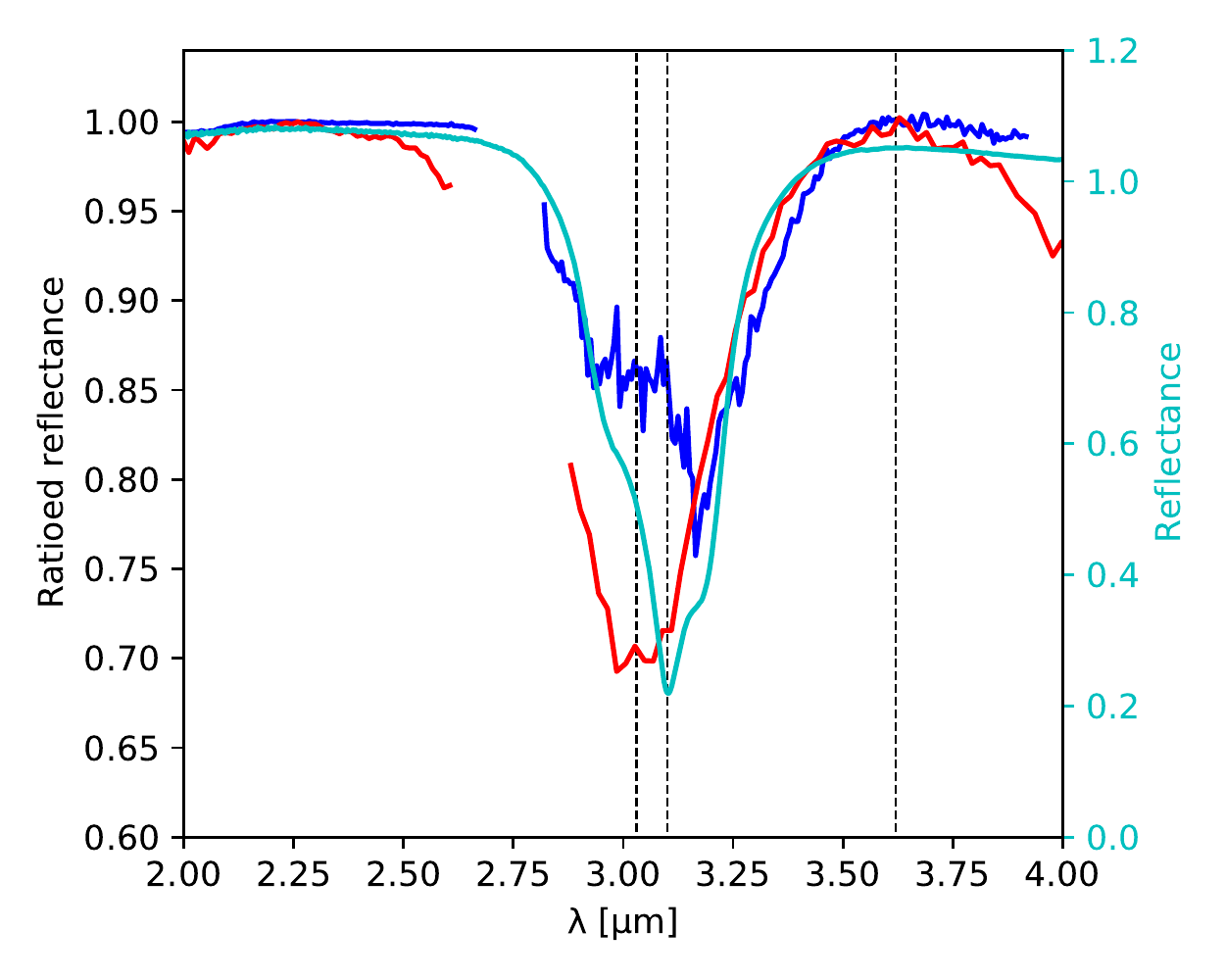}
        \caption{Comparison between the typical spectral ratio associated with the 3~\mum\ northern
            ring identified in this study (red, see \autoref{fig:spectres_typiques} for
            details) and typical small-grained ice transmission spectra. A spectral ratio of a
            Martian water ice cloud obtained with CRISM is shown in dark blue (adapted from
            \citet{vincendon_2011}), and a spectrum of a 0.74~\mum\ thick film of crystalline water
            ice measured under a temperature of 145K (cyan) from \citet{trotta_1996}, available on
            SSHADE \citep{schmitt_2018, schmitt_1996}, is shown in light blue.
            We observe that the position of the absorption band differs between the spectra: about
            3.03~\mum\ for the narrow 3~\mum\ BD, and between 3.1 and 3.2~\mum\ for the
            small-grained atmospheric or surface water ice.  In addition, such small-grained water
            ice with no weak signatures at 1.5 and 2~\mum\ (see also \cite{vincendon_2011}) does not
            present any 4~\mum\ absorption either, unlike our red spectrum.
            Black dotted lines correspond to wavelengths of 3.03, 3.1, and 3.62~\mum.}
        \label{fig:comp_sp_ice}
    \end{figure}

    While water vapor and atmospheric dust can be confidently discarded, atmospheric water ice can
    create a strong and relatively thin spectral absorption near 3~\mum\ \citep{vincendon_2011,
    stcherbinine_2020a}. A particular attention was therefore paid to remove water ice clouds
    through the combination of the 1.5~\mum\ BD and ICI water ice criteria (see
    \autoref{sec:data_filtering}). Nevertheless, as the 3~\mum\ band is stronger than the 1.5~\mum,
    some thin water ice clouds modifying the 3~\mum\ band could remain unfiltered. We illustrate
    this possible effect on \autoref{fig:comp_sp_ice} with a typical water ice cloud spectral ratio
    derived on Mars with CRISM data \citep{vincendon_2011}. We can see that a 3~\mum\ atmospheric
    water ice signature with a significant depth comes with other fainter features at 1.5~\mum\ and
    2~\mum\ that are not observed over the 3~\mum\ northern ring (\autoref{fig:spectres_typiques}).
    In addition, the 3~\mum\ absorption feature due to water ice clouds differs from the observed
    narrow 3~\mum\ band, as ice absorption peaks at $\sim$~3.1 -- 3.2~\mum. 
    Water ice clouds with smaller particle sizes ($r_\mathrm{eff} \sim 0.1~\upmu$m), such as those
    observed frequently at high altitude \citep{stcherbinine_2020a, luginin_2020, liuzzi_2020}, may
    produce a narrower 3~\mum\ absorption band without significant signatures at 1.5 or 1.9~\mum\
    \citep[][figure 4]{vincendon_2011}. However, such clouds do not have a 4~\mum\ absorption
    feature and their 3~\mum\ absorption is centered on $\sim$~3.1~\mum\ instead of 3.03~\mum.
    Overall, observed spectral properties are thus not easily explained by atmospheric water ice.
    Additionally, as mentioned in \autoref{sec:temporal_stability}, the spatial distribution and
    depth of the narrow 3~\mum\ signature are stable over several mid-afternoon observations taken a
    few degrees of $L_s$ apart (cf \autoref{fig:compare_profiles_ls}) and do not show major
    variation between morning ($\sim$~9~a.m.) and evening ($\sim$~6~p.m.) observations
    (\autoref{sec:temporal_stability}), while water ice clouds could be expected to vary
    \citep{Szantai_2021}.

    \subsection{Exposed surface water ice}      \label{sec:discussion_ice}
    The discovery of a northern polar ring, reported here, is only observed poleward of a clear
    boundary located at $\sim$~68\degree N. Its additional 3~\mum\ band is frequently linked with
    water and presents some first-order similarities with water ice
    (\autoref{fig:spectres_typiques}). As atmospheric water ice has been discarded previously, it is
    worth investigating the potential implication of surface water ice.
    
\begin{figure}
        \centering
        \includegraphics[width=\linewidth]{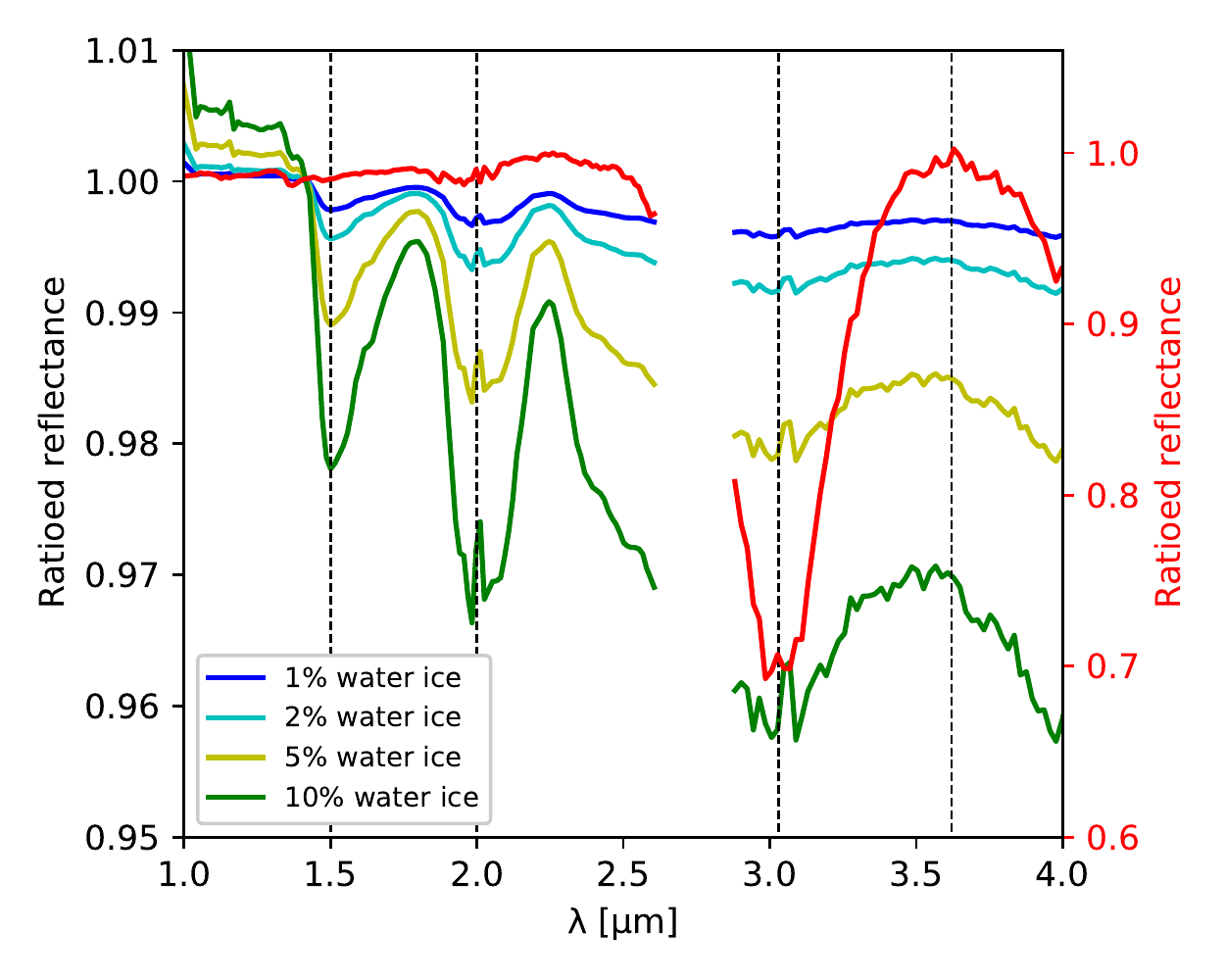}
        \caption{Ratios of spectra simulated with a spatial mix between an OMEGA water ice spectrum
            and a reference spectrum below the 3~\mum\ northern ring ($\sim$ 63\degree N /
            241\degree E) for various amount of water ice, over the same reference spectrum,
            compared to the typical spectral ratio of the 3~\mum\
            northern ring (red, see \autoref{fig:spectres_typiques} for
            details). The water ice spectrum is an average of 9 spectra of the edge of
            the North polar cap (cube 0965\_2, $\sim$ 83\degree N / 133\degree E).
            We observe that even for small amounts of surface water ice within the observed region
            (1 -- 2\%), noticeable 1.5~\mum\ and 2~\mum\ absorption band appears in the spectral
            ratio, along with a blueish spectral slope.
            Black dotted lines correspond to wavelengths of 1.5, 2.0, 3.03, and 3.62~\mum.}
        \label{fig:melange_spatial_waterice}
    \end{figure}

    At such latitudes, the perennial subsurface ice of the permafrost is close to the surface
    \citep{bandfield_2008} and was exposed by the Phoenix robotic arm did \citep{arvidson_2009}
    after digging the ground. However, the permafrost distribution in the polar regions
    \citep[Figure 8]{bandfield_2008} does not match the area detected here, especially between
    $\sim$ 200\degree E and 270\degree E, where the permafrost remains very close to the surface
    down to 50\degree N, while the spectral transition toward the ring-shaped area occurs around
    68\degree N (see \autoref{sec:region_phoenix}). Actually, water ice buried a few cm below the
    surface would be missed by OMEGA as it can only sense the first few micrometers. When some
    subsurface water ice was exposed in a trench dug by the Phoenix robotic arm, observations showed
    that it quickly sublimated \citep{arvidson_2009}, so permafrost is unstable at the surface.
    However, local exposure of pergelisol outcrops have been reported in the northern mid-latitudes
    on pole-facing slopes \citep{dundas_2018, harish_2020, dundas_2021}. These are subpixel compared
    to OMEGA resolution, and we could expect such subsurface ice exposures to be more frequent in
    the northern high latitudes. \cite{dundas_2021} however show that it is not the case
    for the type of exposure they analyzed. Nevertheless, we have tried to simulate the spectral
    signature that OMEGA would have observed by looking at such subpixel permafrost exposures
    (\autoref{fig:melange_spatial_waterice}). We simulated the impact of a spatial mix between an
    observed spectrum of perennial water ice (particles sizes $\sim$ 1 --
    10~\mum) and a typical ice-free high latitude spectrum from \autoref{fig:spectres_typiques}. 
    We observe that even for a few percent of water ice coverage, the spectral ratios exhibit
    1.5~\mum\ and 2~\mum\ absorption bands with intensities comparable to the 3~\mum\ absorption,
    along with a bluer spectral slope in the 1 -- 4~\mum\ range.
    However, neither the blue spectral slope nor sufficient 1.5~\mum\ or 2~\mum\ absorption
    compared to the intensity of the 3~\mum\ feature are present in the 3~\mum\
    northern ring. Overall, these spatial and spectral considerations contradict a subpixel
    permafrost exposures explanation.
    
    Small-grained seasonal surface water ice could also persist throughout summer, e.g. in some
    subpixel shadowed areas. Seasonal water ice is known to be present and stable over shadowed
    pole-facing slopes earlier and later in the season compared to the seasonal cap
    \citep{carrozzo_2009, vincendon_2010}, and some
    surface water ice patches have been observed in the Phoenix region even during the northern
    summer \citep{seelos_2008, cull_2010a}. Additionally, the presence of water ice frost that forms
    during nighttime and remains in the early morning has also been observed by the Viking Lander 2
    Camera 2, the Phoenix lander Surface Stereo Imager (SSI) and CRISM observations of the region
    from $L_s \sim 109^\circ$ \citep{cull_2010a}. Up to $L_s \sim 109^\circ$ the Phoenix landing
    site exposed surface is however ice-free throughout the day \citep{cull_2010a}.  Our
    observations were performed in the mid-afternoon ($\sim$~3~p.m.) from $L_s$ 98\degree\ to
    137\degree, and no surface water ice have been observed during the afternoon hours under Phoenix
    latitudes between $L_s \sim 55^\circ$ and $L_s \sim 156^\circ$ \citep{cull_2010a}, except
    well-known perennial patches of water ice in craters. Additionally, observations of water ice or
    frost in the shadows of pole-facing slopes or craters with CRISM or OMEGA
    \citep{carrozzo_2009, vincendon_2010, harish_2020} exhibit a 1.5~\mum\ absorption of a few
    percent depth along with a stronger 2~\mum\ absorption feature, while we do not detect these
    features. Very small water ice particles are needed to produce a 3~\mum\ band without detectable
    feature at $\lambda = 1.5$~\mum: an optical path length of only 4~\mum\ within ice is supposed
    to be sufficient for OMEGA to detect ice at $\lambda = 1.5$~\mum\ according to
    \cite{vincendon_2010}. We show in \autoref{fig:comp_sp_ice} the spectral properties of a
    0.74~\mum-thick surface water ice  measured at 145K from \citet{schmitt_1996} compared to the
    spectral signature identified here. We observe that even if the width of the two 3~\mum\
    absorption bands are similar, there is a mismatch between their respective positions with a peak
    at 3.03~\mum\ compared to 3.13~\mum\ for the water ice signature. The ice spectrum also lacks
    the 4~\mum\ absorption observed in OMEGA data.
    
    Thus, several issues challenge our attempts to explain the spectral feature of the 3~\mum\
    northern ring with surface ice, either as large- or small-grained.
    
    \subsection{Adsorbed water}     \label{sec:discussion_ads}
    Water adsorption refers to weak bonds between water molecules and a mineral surface, that can
    occur \citep{milliken_2007b} through exchanges with atmospheric vapor \citep{jouglet_2007}. As
    discussed in \autoref{sec:intro}, it has been suggested that adsorbed water may be the main
    explanation for the high latitude increase of the 3~\mum\ band \citep{jouglet_2007, poulet_2008,
    poulet_2010}, although this interpretation has been questioned \citep{audouard_2014b}. 
    
    The increase of the 3~\mum\ spectral feature related to the water adsorption process comes
    generally along with an increase of the 1.9~\mum\ band (\citep[figure 16]{pommerol_2009} and
    \cite{beck_2015}). However, we observe that the narrow 3~\mum\ and the 4~\mum\ bands identified
    in this study over the 3~\mum\ northern ring are anti-correlated with the areas where the
    1.9~\mum\ band is high (\autoref{fig:multi_polar_maps_nord}). Additionally, the spectral
    ratio typical of the detected area does not contain a significant 1.9~\mum\ band
    (\autoref{fig:spectres_typiques}), although the 1.8 -- 2.1~\mum\ range may contain a shallow
    absorption ($<$ 1\%). This does not favor the adsorbed water hypothesis.
    
\begin{figure}[h!]
        \centering
        \includegraphics[width=\linewidth]{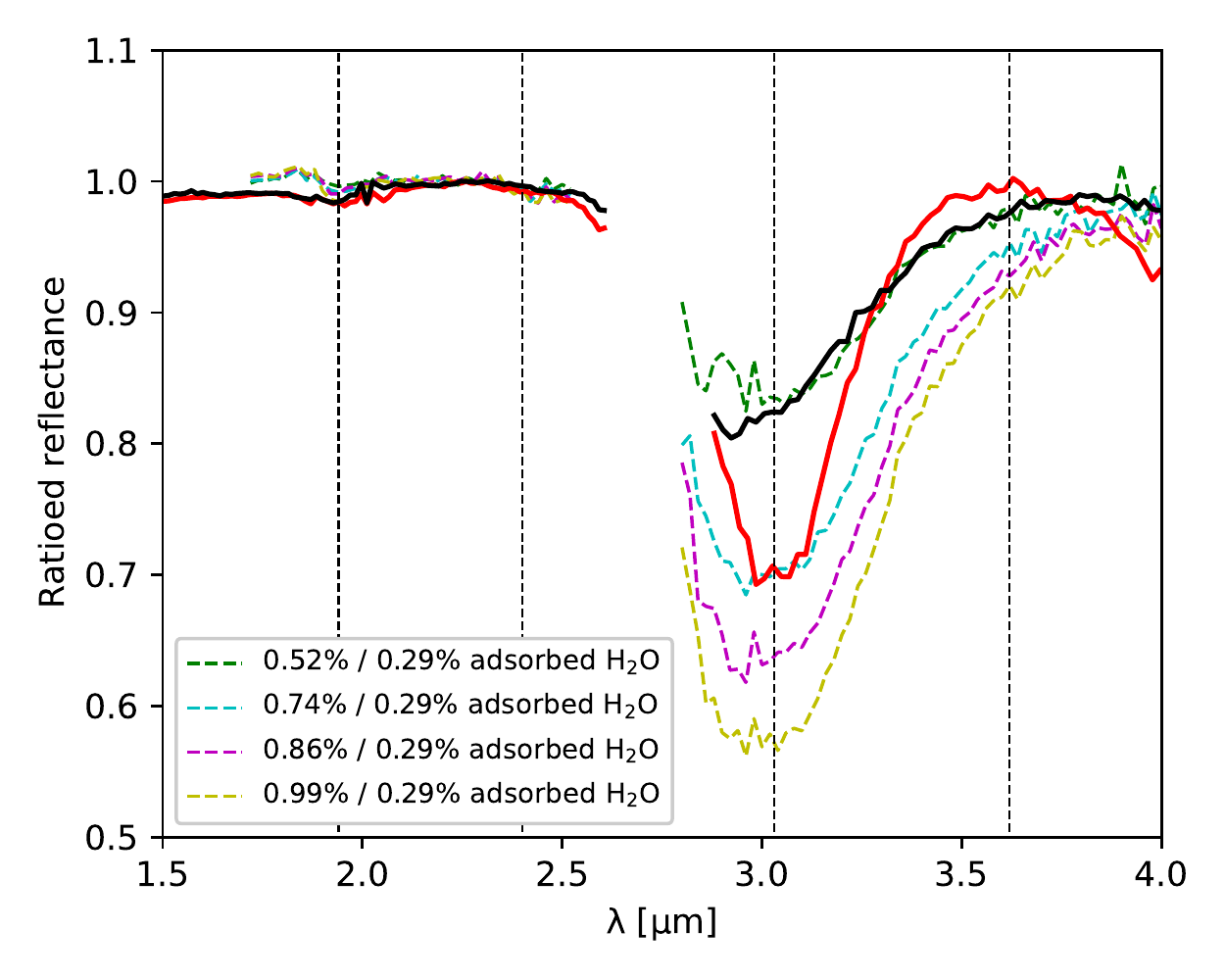}
        \caption{Comparison between high latitude OMEGA spectral ratios and ratios of laboratory
            materials with different amount of adsorbed water. The spectral ratio identified in this
            study over the 3~\mum\ northern ring (red solid line) and the ratio obtained
            outside the 3~\mum\ northern ring between 50\degree N and 60\degree N (black solid line)
            (see \autoref{fig:spectres_typiques} for details) are compared to colored dotted
            spectra corresponding to ratios of spectra acquired on the same Volcanic tuff sample at
            different amounts of adsorbed H$_2$O from \citet{pommerol_2009} and available on SSHADE
            \citep{schmitt_2018, pommerol_2007}.
            Black dotted lines correspond to wavelengths of 1.94, 2.4, 3.03, and 3.62~\mum.}
        \label{fig:comp_sp_adsorption}
    \end{figure}

    We compare in \autoref{fig:comp_sp_adsorption} the spectral ratio of the 3~\mum\ northern
    ring to spectral ratios of a Volcanic tuff sample with different levels of hydration from
    \citet{pommerol_2009}.  This sample was chosen because it does not present a strong absorption
    at 1.9~\mum, but a variation of the 3~\mum\ band. We have however checked that the spectral
    behavior of adsorbed water estimated with this sample was typical of other samples presented in
    \citep{pommerol_2009}. We can see that the adsorbed water spectral shape does not compare well
    with the spectral ratio of the 3~\mum\ northern ring. Indeed, we can see that the 3~\mum\ band
    associated with water adsorption is significantly wider than the narrow 3~\mum\ absorption and
    that the adsorption spectra are unable to reproduce the decrease of reflectance between 3.7 and
    4~\mum. The 3~\mum\ northern ring spectral signature thus seems to be related to another
    process, and adsorbed water does not seem to be the main explanation for the 3~\mum\ band
    increase observed poleward 68\degree N.
    
    On the other hand, we observe that the spectral ratio obtained southward of the 3~\mum\ northern
    ring, that is between 50\degree N and 60\degree N, provides a reasonable 3~\mum\ shape match
    with adsorbed water (\autoref{fig:comp_sp_adsorption}), along with a small 1.9~\mum\ band
    that is also consistent with adsorbed water (\autoref{fig:spectres_typiques}). The
    observed spectral evolution between 50\degree N and 60\degree may thus reveal an increase in the
    amount of adsorbed water south of the ring. However, the observed minimum of the 3~\mum\ band is
    at $\sim$~2.9~\mum\ in the observed ratio, which is not comparable to adsorbed water.
    Additionally, as discussed previously (\autoref{sec:intro}), adsorbed water is difficult to
    reconcile with the temperatures required to release the water from the soil measured by Phoenix
    \citep{smith_2009a} and the lack of temporal variability of the  3~\mum\ band
    \citep{audouard_2014b}. The fact that such an increase of adsorbed water would not be observed
    at the higher latitudes of the ring is also puzzling. Alternatively, an increasing amount of
    hydrated minerals with latitude would also result in such a spectral ratio with 3~\mum\ and
    1.9~\mum\ bands of comparable shapes \citep{jouglet_2007, beck_2015}. The spectral ratio
    obtained southward of the 3~\mum\ northern ring may thus indicate the presence of an enhanced
    hydration within minerals or amorphous phases south of the ring.
    
    \subsection{Surface salts}      \label{sec:discussion_mineral}
    As discussed in previous sections, the spectral signature identified over the 3~\mum\ northern
    ring does not seem to be associated with atmospheric, icy or adsorbed water phenomena.
    Consequently, it must relate to some sort of modification of the physical and/or chemical
    properties of the surface. 
    
    Studies of the mineralogy of the north polar regions have revealed that the highest amounts of
    hydrated minerals detectable in the near-IR seem to be located in the low-albedo regions
    surrounding the perennial polar cap \citep{langevin_2005, poulet_2008, masse_2010, masse_2012}.
    This area also presents a high density of dune fields \citep{tanaka_2008, masse_2012}. These
    polar hydrated minerals have been mostly identified as Ca-sulfates, and more specifically gypsum
    \citep{langevin_2005, horgan_2009, masse_2010, masse_2012}. The highest concentration is located
    in the Olympia Undae deposit \citep{langevin_2005}. Perchlorates are an alternative candidate
    for the spectral shapes typical of these latitudes \citep{masse_2010, hanley_2015}. While these
    low albedo terrains are not included in the area of the 3~\mum\ northern ring (see previous
    discussion), this indicates that sulfates, in particular Ca-sulfates, and possibly perchlorates,
    are abundant and widespread in the northern polar regions. The Phoenix lander is on the contrary
    located within the 3~\mum\ northern ring. Phoenix also provided indication that sulfates
    (notably Mg-sulfates) and, with lower concentrations, perchlorates, participate in the
    composition of high altitude soils with typical concentrations of a few weight \%
    \citep{boynton_2009, hecht_2009, smith_2009a, toner_2014}.
    
\begin{figure}[h!]
        \centering
        \includegraphics[width=\linewidth]{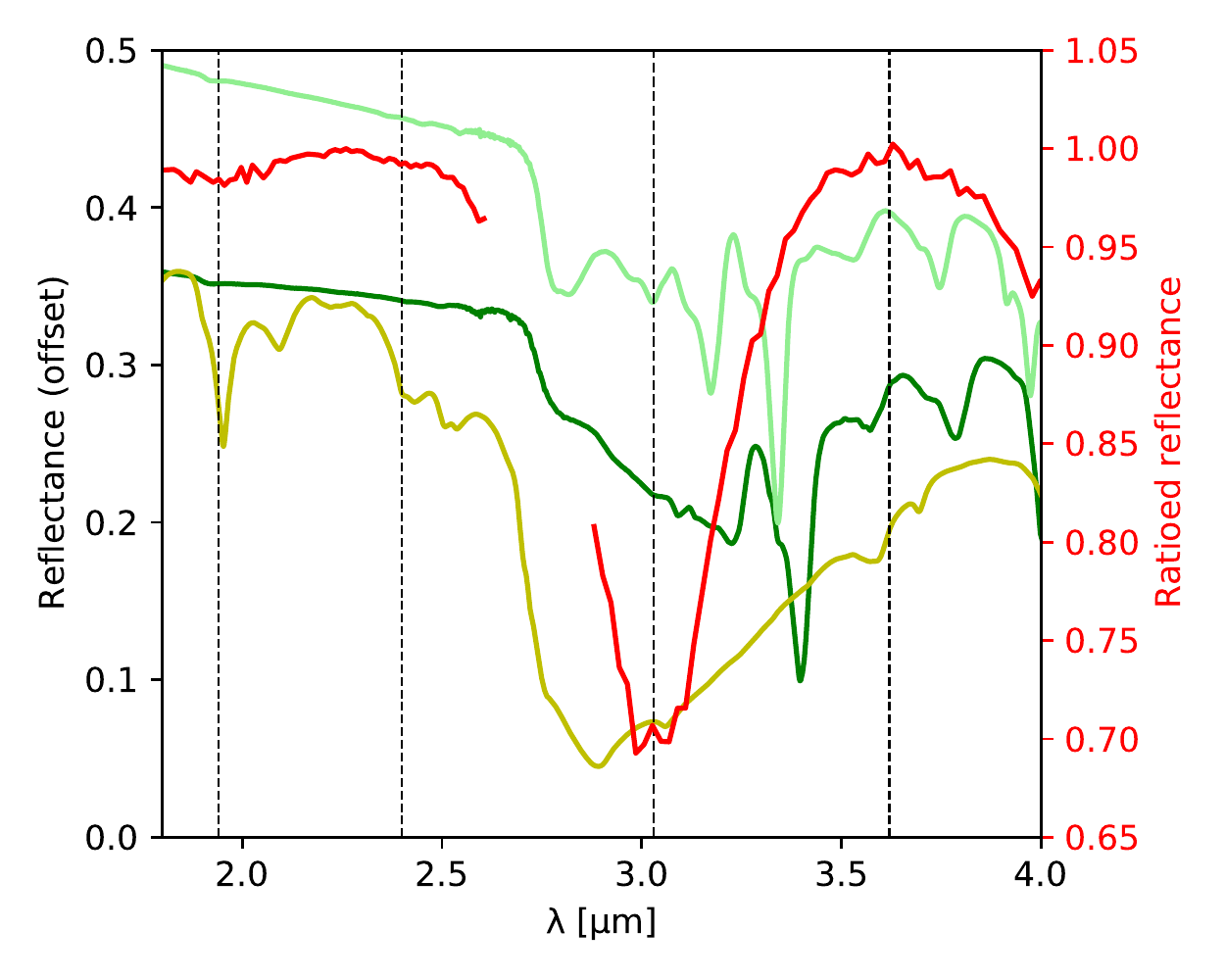}
        \caption{Comparison between the typical spectral ratio associated with the 3~\mum\
            northern ring  band (red, see \autoref{fig:spectres_typiques} for
            details) with laboratory spectra of various perchlorates: synthetic anhydrous samples of
            KClO$_4$ (dark green, JB983, rescaled by a factor 3 and offset for clarity) and
            NaClO$_4$ (light green, JB982, rescaled by a factor 3 and offset for clarity); and a
            natural salt from the Atacama desert in Chile in yellow (JBE47), with particles sizes
            lower than 125~\mum. Black dotted lines correspond to wavelengths of 1.94, 2.9,
            3.03, and 3.62~\mum. RELAB spectra downloaded from the PDS spectral library
            \citep{milliken_2020}.}
        \label{fig:comp_sp_perchlorates_ini}
    \end{figure}

    We first compare our signature with spectra from various perchlorate salts in 
    \autoref{fig:comp_sp_perchlorates_ini}. Anhydrous samples are featureless at
    wavelengths lower than 2.6~\mum, similarly to the 3~\mum\ northern ring spectral
    signature, but with a very different spectral shape at longer wavelengths devoid of a marked
    3~\mum\ band. On the other hand, hydrated perchlorate salts do have a strong 3~\mum\
    band but centered at 2.8-2.9~\mum\ and accompanied with other features notably at shorter
    wavelengths, which is again poorly compatible with the 3~\mum\ northern ring spectral
    signature.
    
\begin{figure}[h!]
        \centering
        \includegraphics[width=\linewidth]{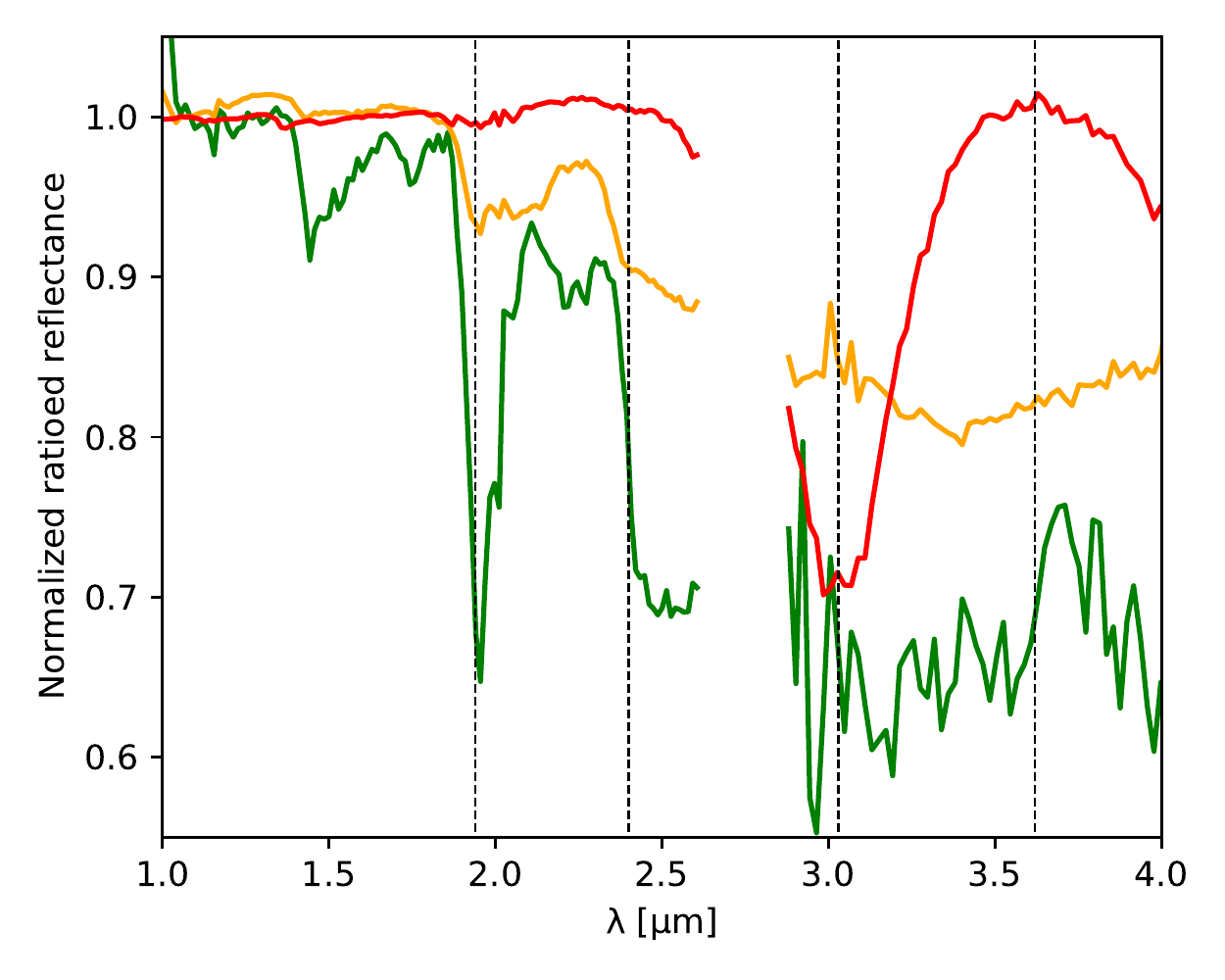}
        \caption{Comparison between the typical spectral ratio associated with the 3~\mum\
            northern ring (red, see \autoref{fig:spectres_typiques}\ for
            details) and ratios obtained over areas previously identified by OMEGA as sulfates-rich
            at Terra Meridiani (orange, $\sim$ 1.7\degree N / 0.1\degree W over $\sim$ 0.5\degree N
            / 0.4\degree E) and in the north polar region of Olympia Undae (green, $\sim$
            80.1\degree N / 244.1\degree E over $\sim$ 77.4\degree N / 68.9\degree E).
            The ring spectrum is not comparable to previously reported OMEGA sulfates spectra. Black
            dotted lines correspond to wavelengths of 1.94, 2.9, 3.03, and 3.62~\mum}.
        \label{fig:comp_sp_sulfates_omega}
    \end{figure}

    The observed 4~\mum\ absorption in our OMEGA spectra shows some similarities with the bluer wing
    of the wide 4.4 -- 4.7~\mum\ absorption of sulfates due to SO$_4$ vibrations harmonic
    \citep{bishop_2014, bishop_2019a}. 
    We compare in \autoref{fig:comp_sp_sulfates_omega} the spectral signature of
    the 3~\mum\ northern ring with typical OMEGA spectra of previously identified sulfates-rich
    areas: Olympia Undae \citep{langevin_2005a} and Terra Meridiani
    \citep{gendrin_2005a}. These hydrated sulfates are associated with strong 1.9 and
    2.4~\mum\ absorption bands not clearly observed in the 3~\mum\ northern ring spectrum.
    They also provide a poor fit of the 3~\mum\ band, with a wider shape and a band center
    shifted toward longer wavelengths.

\begin{figure}[h!]
        \centering
        \includegraphics[width=\linewidth]{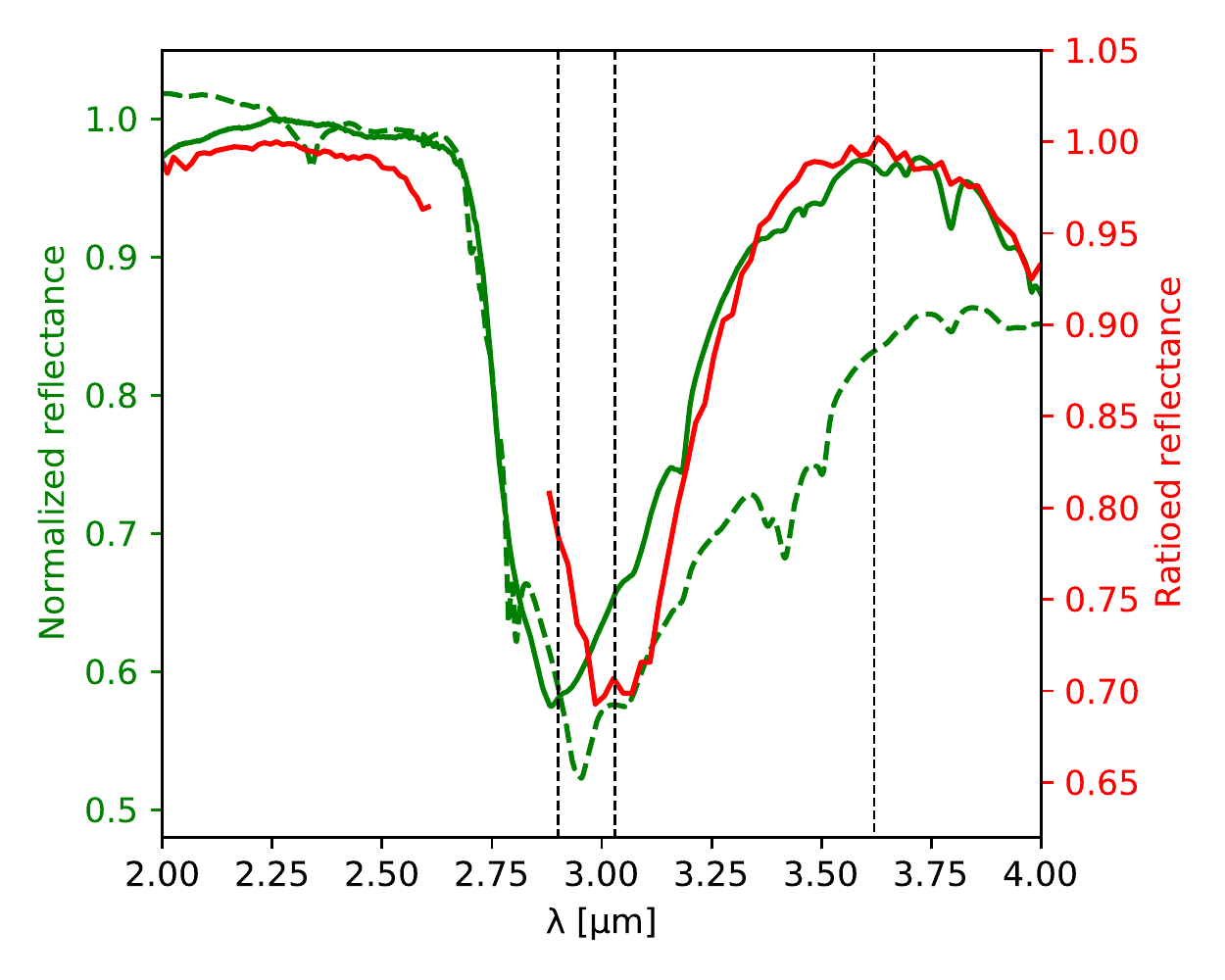}
        \caption{Comparison between the typical spectral ratio associated
            with the 3~\mum\ northern ring (red, see \autoref{fig:spectres_typiques} for
            details) and RELAB spectra (green) of Ca-sulfate anhydrite (solid line: GDS42; dashed
            line: JB641A). 
            The GDS42 anhydrite provide a satisfactory fit of the 4~\mum\ absorption, the
            lack of absorption in the $\sim$~2~\mum\ spectral range, and the narrowness of the
            3~\mum\ band, but with a band minimum at $\sim$~2.9~\mum\ that is not
            comparable to the ring. The JB641A anhydrite does not have a 4~\mum\ feature nor a
            narrow 3~\mum\ band, but its 3~\mum\ band is centered on
            $\lambda~\sim$~3.0~\mum\ similarly to the ring. Black dotted lines correspond to
            wavelengths of 2.9, 3.03, and 3.62~\mum. RELAB spectra downloaded from the USGS
            spectral library \citep{kokaly_2017} and PDS spectral library
            \citep{milliken_2020}.}
        \label{fig:comp_sp_anhydrite}
    \end{figure}

    The lack of any significant signature in the 1 -- 2.5~\mum\ spectral range may 
    suggest a weakly-hydrated sulfate \citep{bishop_2014}. Such sulfates are known to be present on
    Mars since a small amount of anhydrite has been detected in Gale Crater by MSL
    \citep{bish_2013}. We compare the spectrum identified in this study over the 3~\mum\ northern
    ring to anhydrite in \autoref{fig:comp_sp_anhydrite}. The 4~\mum\ band
    shape, the relative depth of the 4 and 3~\mum\ bands, and the narrow 3~\mum\ band width can be
    reproduced by the anhydrite sample. However, the 3~\mum\ band minimum in this anhydrite sulfate
    sample is not located at the same wavelength as in our observed spectrum. The spectrum of the
    3~\mum\ northern ring is also compared to another anhydrite sample that contains minor
    admixtures \citep{bishop_2014}: this lowers the overall quality of the fit, but we can see that
    a subtle contamination of this sulfate by other mineral components shifts the minimum of the
    3~\mum\ band to longer wavelengths, bringing it closer to our observation.

    Studies of the distribution of gypsum across the north circumpolar dune field showed evidence
    for eolian processes that may have scattered the gypsum away from a few dune field reservoirs
    \citep{horgan_2009, masse_2012}. The smallest grains could have been transported further away,
    and could have accumulated in the bright area, as bright areas on Mars usually correspond to
    areas where erosion of fines and dust is less efficient. Exposed to southern, warmer conditions,
    gypsum grains could have undergone desiccation toward bassanite and then even to anhydrite
    \citep{poitras_2018}. Alternatively, such desiccation could have occurred at the source and
    could have then facilitated fracturing, producing smaller and wind transportable grains. These
    desiccated sulfate grains could have been transported further away by winds compared to larger
    gypsum grains.

    \begin{figure}[h!]
        \centering
        \includegraphics[width=\linewidth]{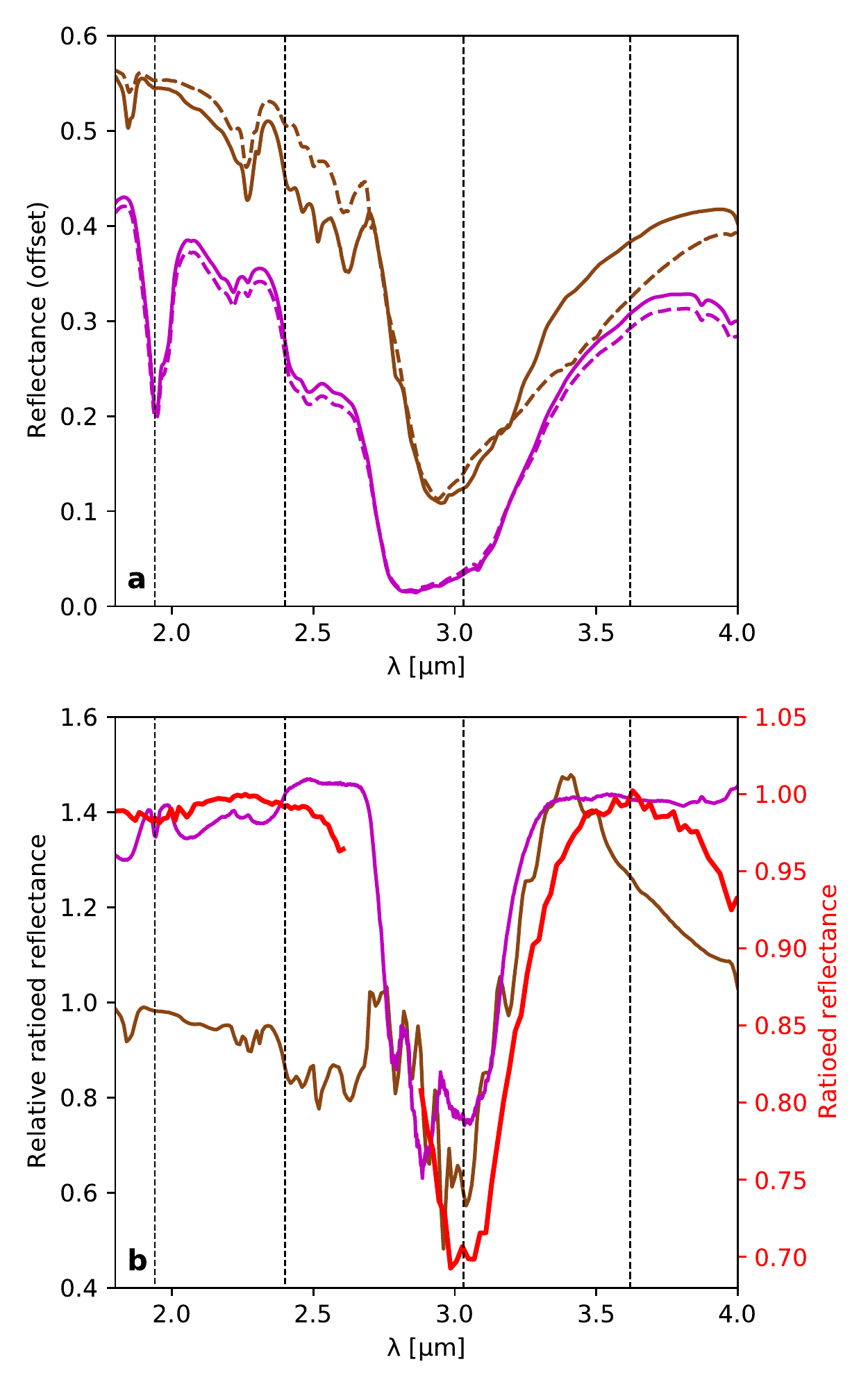}
        \caption{(a) RELAB spectra of Jarosite (brown, offset of 0.1 for clarity)
            from two different places: Rublev (solid line, JBA78) and St Leger (dashed line, JB701)
            ; and spectra of Gypsum (purple, JBE64) for two grain size: $<$ 45~\mum\ (solid line)
            and 45--90~\mum\ (dashed line). (b) Brown: ratio of jarosites (Rublev over St Leger) ;
            Purple: ratio of gypsum ($<$~45~\mum\ over 45--90~\mum\, rescaled by a factor 4 for
            clarity); Red: spectral ratio associated with the 3~\mum\ northern ring. Both sulfates
            ratios exhibit a narrow 3~\mum\ band that can be compared to that of the ring, along
            with an absorption at 4~\mum\ for the brown spectrum.
            Black dotted lines correspond to wavelengths of 1.94, 2.4, 3.03, and 3.62~\mum.
            RELAB spectra downloaded from the PDS spectral library \citep{milliken_2020}.}
        \label{fig:comp_sp_jar}
    \end{figure}

    We compare in \autoref{fig:comp_sp_jar} two gypsum spectra obtained for two different grain
    sizes. While the gypsum spectrum in itself provides a poor fit to our observed spectrum, the
    ratio of the small grain size over the larger grain size provides a relatively satisfactory
    explanation for the narrow 3~\mum\ band of the ring. This may indicate that reducing the grain
    size of calcium-sulfates can produce the spectral evolution observed at the 3~\mum\ northern
    ring. The 4~\mum\ feature is however absent of the gypsum ratio of two sizes, so reducing the
    size does not seem to be enough. Additionally, this interpretation would require calcium
    sulfates with larger grains to be present southward of the identified area, as our spectrum is a
    ratio between 70\degree N and 63\degree N, while observational evidence for gypsum have been
    obtained essentially poleward. Southward, as discussed in \autoref{sec:discussion_ads}, terrains
    located between $\sim$ 50\degree N and $\sim$ 65\degree N do show enhanced hydration spectral
    features at 1.9~\mum\ and 3~\mum\ that may be related to hydrated phases such as hydrated
    minerals. This could indicate that soils south of the 3~\mum\ northern ring actually contain
    more altered phases compared to equatorial latitudes, as previously proposed (see
    \autoref{sec:intro}). The 3~\mum\ northern ring would then correspond to a latitude transition
    beyond which these phases are modified.
    
    While a sole modification of the grain size does not seem to explain all observed features
    (notably, the 4~\mum\ band, see previous paragraph), one could imagine that, if low amount of
    sulfates are already present south of the 3~\mum\ northern ring, increasing the latitude can
    modify the hydration level of sulfates and produce the observed spectral features in ratios. To
    test this hypothesis, we compare in \autoref{fig:comp_sp_jar}b our observation to a ratio of two
    samples of jarosite from two different locations on Earth. Jarosite is a hydrous
    sulfate of potassium and ferric iron detected on Mars by the Opportunity rover
    \citep{klingelhofer_2004} which may form under icy conditions
    \citep{niles_2020}. Both samples
    are described as jarosite mixture with similar grain sizes lower than 45~\mum. One of the two
    samples appears more hydrated from the 1.9~\mum\ band depth. Ratioing the spectra
    of these two samples results in a spectral shape that presents some similarities with our
    observation, with both a narrow 3~\mum\ band centered at the appropriate wavelength and a
    decrease toward 4~\mum. This example illustrates that increasing the amount of hydration of
    sulfates may result in a spectral ratio consistent with our observations. Actually, there
    are indications that soils contain minor amounts of anhydrous sulfates (anhydrite) even at
    equatorial latitudes \citep{bish_2013}, so trace amount of lowly
    hydrated sulfates may already be present southward of the ring.
    
     We also computed spectral ratios between the perchlorates samples discussed previously in
     \autoref{fig:comp_sp_perchlorates_ini}. While none of these ratios provide a
     satisfactory explanation on his own, they illustrate again that ratioing salts spectra from
     different samples may result in 3~\mum\ bands that share some of the characteristics
     (narrowness or band center) of the 3~\mum\ northern ring. As indicated by Phoenix
     measurements, the salts content of soil at these latitudes probably results from a mixture of
     various species \citep{boynton_2009, hecht_2009, smith_2009a, toner_2014}, which
     could explain why none of the spectra or ratios used here for comparison perfectly explain the
     observed spectral shape, as they are each inherited from relatively pure samples. The
     3~\mum\ northern ring may then result from a modification of the hydration of a mixture of
     various salts contaminants in the soils.
    
     \begin{figure}[h!]
        \centering
        \includegraphics[width=\linewidth]{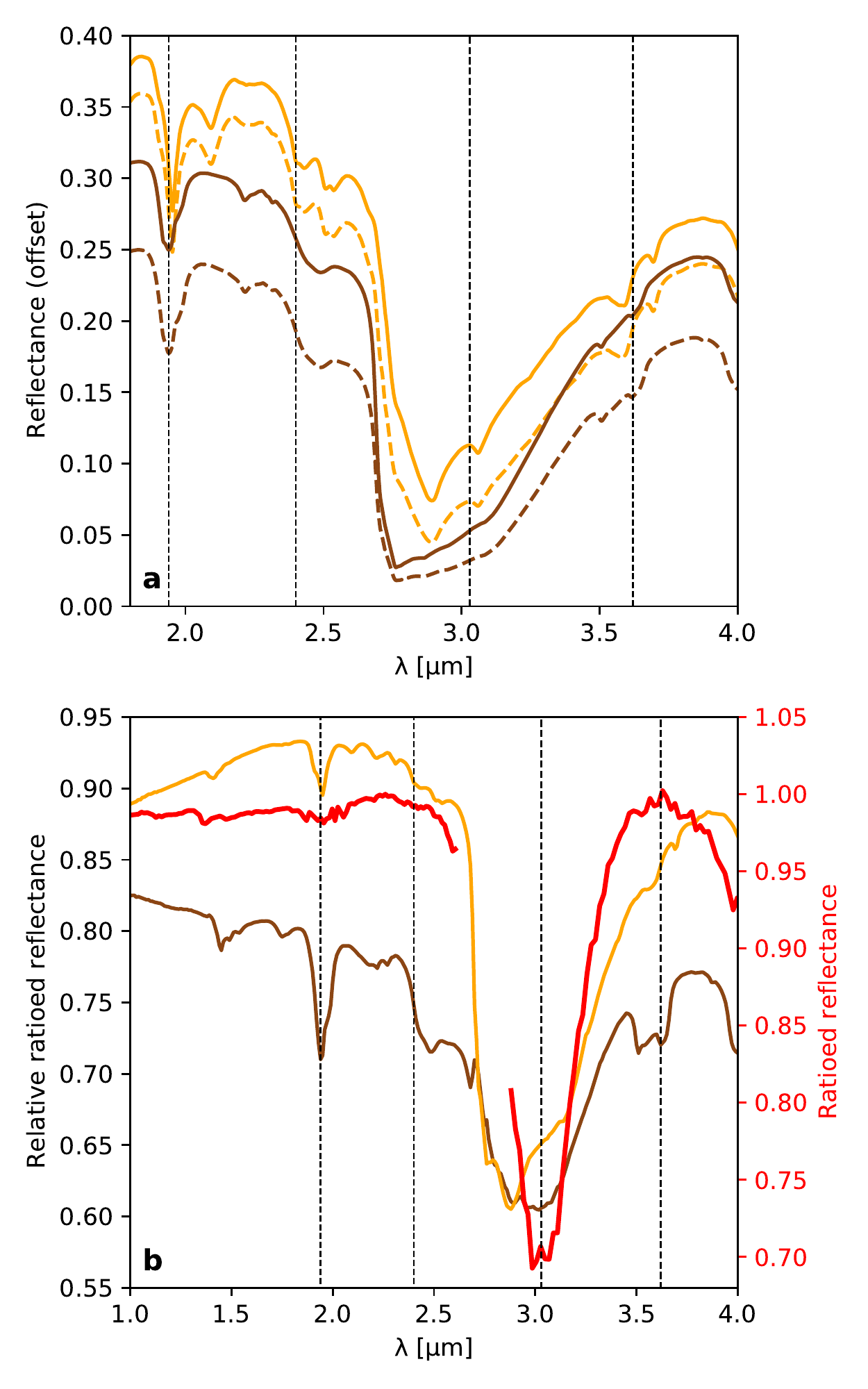}
        \caption{(a) RELAB spectra of natural perchlorate salts from the Atacama desert in
            Chile. In orange, two different samples with particles sizes lower than 125~\mum\
            (dashed line JBE47, solid line JBE48). In brown, a third sample (JBE49) for two ranges
            of particle sizes: lower than 125~\mum\ (solid line), and between 125 and
            250~\mum\ (dashed line). (b) Comparison between the typical spectral ratio
            associated with the 3~\mum\ northern ring  band (red, see
            \autoref{fig:spectres_typiques} for details) and perchlorate salts ratios calculated
            using the laboratory spectra shown in panel (a). Orange: ratio of two natural samples
            with particles sizes lower than 125~\mum\ (JBE47 / JBE48); the presence of the 1.9~\mum\
            band in the ratio may indicate an hydration level difference between both. Brown: ratio
            of the 125--250~\mum\ over the $<$~125~\mum\ size for the JBE49 sample. Black dotted
            lines correspond to wavelengths of 1.94, 2.9, 3.03, and 3.62~\mum.
            RELAB spectra downloaded from the PDS spectral library \citep{milliken_2020}.}
        \label{fig:comp_sp_perchlorates}
    \end{figure}

\begin{figure}
        \centering
        \includegraphics[width=\linewidth]{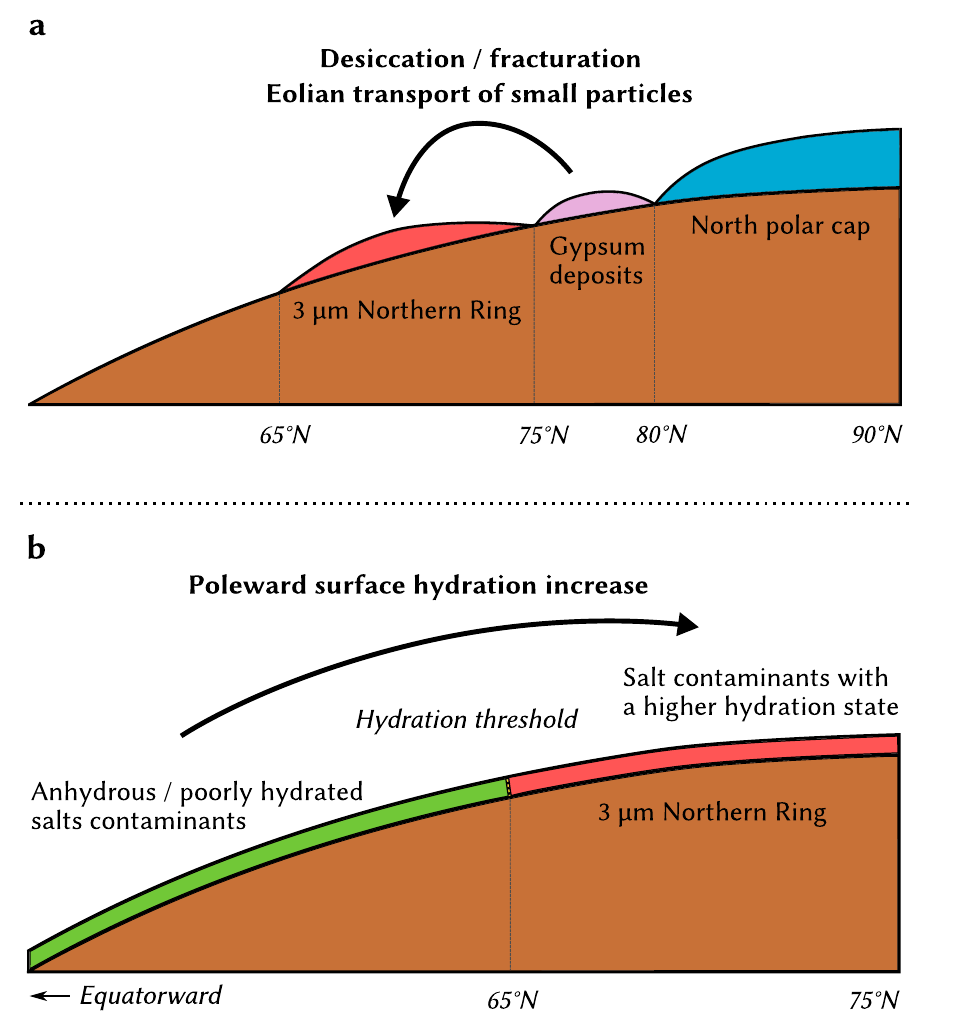}
        \caption{Schematic drawings of two possibles scenarios for the formation of the
            3~\mum\ northern ring. 
            (a) Hydrated salts from the northern regions are transported to lower latitudes through
            eolian processes with a possible contribution of fracturation and desiccation either at
            source or arrival site. 
            (b) A $\sim$~65\degree N latitude threshold separates two hydration states of salts
            (sulfates and/or perchlorates) contaminants within soils, with more hydration over the
            3~\mum\ northern ring.}
        \label{fig:schemas_scenarios}
    \end{figure}

    Overall, while the exact mechanism responsible for the spectral signature of the ring is not
    firmly identified, available constraints point toward a possible contribution of salts,
    notably sulfates, but also possibly perchlorates, in a context of hydration state change
    and/or grain size change. We summarize the putative formation scenarios mentioned in the
    previous paragraphs in \autoref{fig:schemas_scenarios}. As our observed 3~\mum\
    northern ring is superficial and overlays different geological units dated from early-Amazonian
    (cf \autoref{sec:north_regions}), we are probably observing a relatively recent process. This
    process may be currently acting and somehow related to the seasonal ice cap
    \citep{audouard_2014b} that cover the area every winter, although this association is not clear
    as the cap extent at southern latitudes, down to near 50\degree N \citep{appere_2011}.
    Alternatively, this modification of the surface may be related to a longer timescale. Mars high
    latitudes are currently covered by a mantle that is thought to have been deposited during past
    glaciations over the last million years \citep{head_2003}. In both hemispheres, this ice-rich
    mantle is considered to be present and uniform poleward $\sim$ 60\degree, while being only
    partially preserved equatorward \citep{head_2003}. Sulfates formation can be associated with ice
    on Earth and this has been proposed as an explanation for sulfates formation on Mars at high
    northern latitudes \citep{masse_2010} or near the equator \citep{niles_2009}, either through
    authigenic formation within ice or associated with atmospheric ice condensation and
    precipitation \citep{masse_2010}. Widespread spectral features possibly related to sulfates have
    been observed at high southern latitudes \citep{poulet_2008, wray_2009, ackiss_2014,
    carter_2016}, and widespread spectral signatures potentially linked with limited water
    alteration have been observed at high northern latitudes in this study (see
    \autoref{sec:discussion_ads} and previous paragraph) and previously \citep{wyatt_2004,
    michalski_2005, horgan_2012}. The 3~\mum\ northern ring identified in this study north of $\sim$
    65\degree\ N may then highlight a modification of such a widespread sulfates contamination. This
    modification could be a hydration state increase occurring once a latitude threshold is crossed
    in the northern hemisphere where the amount of water vapor and the overall atmospheric pressure
    are higher (see \autoref{fig:schemas_scenarios}b).
    
\section{Conclusion}
    In this paper, we report and characterize the detection of a new spectral feature defined by a
    deep and relatively narrow 3~\mum\ band, centered at 3.03~\mum, that differs from the usual Mars
    3~\mum\ band, coupled with a shallow but significant wide 4~\mum\ feature. This spectral
    signature appears to be specific to the bright regions located at high northern latitudes. More
    precisely, the detection occurs over a wide, open ring area surrounding the cap between
    $\sim$~68\degree N and $\sim$~76\degree N and extending from $\sim$~0\degree E to
    $\sim$~315\degree E. This area includes the Phoenix landing site and is referred as the "3~\mum\
    northern ring" in this study.
  
    A detailed study of the spectral properties observed in this ring (3~\mum\ band shape,
    association with other spectral features, time variability) leads to rule out atmospheric and
    water-ice explanations. The spectral properties cannot be explained by an
    increase in the amount of weakly bound adsorbed water either.
    
    The ring appears to possess a specific surface signature that may relate to some sort of
    modification of the physical or chemical properties of the ice-free soils. The ring occurs north
    of terrains where a progressive latitudinal increase of surface hydration (1.9~\mum\ and
    3.0~\mum\ bands) is observed. The spectral signature of the ring does however not correlate with
    signatures in the 1 -- 2.5~\mum\ spectral range, despite the presence of specific hydrated
    signatures at similar latitudes in nearby dark terrains.
    
    First comparisons with spectra from libraries suggest that salts, notably
    sulfates, and possibly perchlorates, may play a role.
    The most promising comparisons are obtained with the anhydrous
    Cal-sulfate anhydrite or with ratios of sulfates and/or perchlorates spectra suggestive of an
    hydration or a grain size modification. In this context we propose two hypothetical scenarios to
    explain this observation. This first one involves aeolian transport and dessication of the
    gypsum deposits previously identified near the north polar cap. The second relies on the
    presence of trace amounts of salts within soils which hydration level is modified on each side
    of a latitudinal threshold at $\sim$~65\degree N. Within both frameworks, the 3~\mum\ northern
    ring could highlight ice-related formation or modification pathways of salts on Mars.


\section*{Data availability}
The OMEGA/MEx data are freely available on the ESA PSA at 
\url{https://archives.esac.esa.int/psa/#!Table%20View/OMEGA=instrument}.
The source code of the \emph{OMEGA-Py} Python module is freely available on
GitHub at \url{https://github.com/AStcherbinine/omegapy}. 
Laboratories spectra were downloaded from the USGS spectral library
\citep{kokaly_2017}, the PDS
spectral library \citep{milliken_2020} and SSHADE \citep{schmitt_2018}.

\section*{Acknowledgments}
We thank A. Pommerol and B. Bultel for their constructive reviews that improve the quality of the
manuscript.

\printcredits

\bibliographystyle{cas-model2-names}

\bibliography{bibliography}

\end{document}